\documentclass[prd,aps,twocolumn,superscriptaddress,amsmath,amssymb,nofootinbib]{revtex4-2}
\usepackage{epsfig}
\usepackage{graphicx}
\usepackage[dvipsnames]{xcolor}
\usepackage{upgreek}
\usepackage{hyperref}
\hypersetup{
	colorlinks=true,
	pdfborder={0 0 0},
	citecolor=blue,
	linkcolor=blue,
	urlcolor=blue
}
\usepackage[normalem]{ulem}

\usepackage{comment} 

\usepackage{tikz}
\usetikzlibrary{quantikz}

\begin{document}

\title{Phase-Space methods for neutrino oscillations: extension to multi-beams} 

\author{Denis Lacroix } \email{lacroix@ijclab.in2p3.fr}
\affiliation{Universit\'e Paris-Saclay, CNRS/IN2P3, IJCLab, 91405 Orsay, France}
\author{Angel Bauge} 
\affiliation{Universit\'e Paris-Saclay, CNRS/IN2P3, IJCLab, 91405 Orsay, France}
\author{Bulent Yilmaz}
\affiliation{Physics Department, Faculty of Sciences, Ankara University, 06100 Ankara, Turkey}
\author{Mariane Mangin-Brinet} 
\affiliation{Laboratoire de Physique Subatomique et de Cosmologie, CNRS/IN2P3, 38026 Grenoble, France}
\author{Alessandro Roggero}
\affiliation{Dipartimento di Fisica, University of Trento, via Sommarive 14, I–38123, Povo, Trento, Italy}
\affiliation{INFN-TIFPA Trento Institute of Fundamental Physics and Applications,  Trento, Italy}
\author{A. Baha Balantekin}
\affiliation{Department of Physics, University of Wisconsin--Madison,
Madison, Wisconsin 53706, USA}

\date{\today}
\begin{abstract}
The Phase-Space approach (PSA), which was originally introduced in [Lacroix et al., Phys. Rev. D 106, 123006 (2022)] to describe neutrino flavor oscillations for interacting neutrinos emitted from stellar objects is extended to describe arbitrary numbers of neutrino beams. The PSA is based on mapping the quantum fluctuations into a statistical treatment by sampling initial conditions followed by independent mean-field evolution. A new method is proposed to perform this sampling that allows treating an arbitrary number of neutrinos in each neutrino beams. We validate the technique successfully and confirm its predictive power on several examples where a reference exact calculation is possible. We show that it can describe many-body effects, such as entanglement and dissipation induced by the interaction between neutrinos. Due to the complexity of the problem, exact solutions can only be calculated for rather limited cases, with a limited number of beams and/or neutrinos in each beam. The PSA approach considerably reduces the numerical cost and provides 
an efficient technique to accurately simulate arbitrary numbers of beams. Examples of PSA results are given here, including up to 200 beams with time-independent or time-dependent Hamiltonian. We anticipate that this approach will be useful to bridge exact microscopic techniques with more traditional transport theories used in neutrino oscillations. It will also provide important reference calculations 
for future quantum computer applications where other techniques are not applicable to classical computers.    
\end{abstract}


\maketitle

\section{Introduction and notations}

Neutrinos emitted during astrophysical processes such as the core-collapse supernovae or the neutron star mergers 
offer an important source of information on physical phenomena occurring inside dense stellar objects \cite{Dua06,Bah07,Vol24}. Once emitted, high neutrino flux might interact with matter through 
the Mikhe\"iev-Smirnov-Wolfenstein (MSW) effect \cite{Wol78,Mik85}. Neutrinos themselves can interact with each other \cite{Ful87,Not88,Sig93}. The problems of neutrinos escaping from stellar emitters 
correspond to many particles interacting at the early stage with the surrounding matter. These effects tend to influence the highly coherent flavor oscillation process by inducing entanglement between particles. Due to the large number of particles involved, neutrino propagation is usually treated through hydrodynamical or semiclassical transport theories \cite{Dua06,Vol24}. 

Alternatively, by grouping neutrinos in beams and by making simplifying assumptions on 
their interaction \cite{Peh11}, one can also attack the problem from a many-body perspective and try 
to solve the Hamiltonian dynamics exactly by full configuration interaction techniques. This is usually done by neglecting the MSW effects and focusing on neutrino-neutrino interactions \cite{Peh11,Bir18,Pat19,Rra19,Mar21,Xio22,Ill22,Lac22,Cer22,Rog22a,Mar23a,Mar23b,Bha23}. These brute-force approaches are useful to study quantum effects like entanglement between neutrinos \cite{Cer19,Rog21,Pat21,Bal22}. This strategy brings new information compared to other transport theories but becomes 
extremely challenging for classical computers when the number of beams or neutrinos per beam increases. 

In recent years, an effort has been made to apply the full configuration interaction approach to quantum computers \cite{Hal21,Yet22,Kum22,Ill22b,Ami23,Tur24}, and the neutrino flavors oscillations problem has emerged 
as possible early pilot applications for quantum technologies. In the present work, we briefly discuss the quantum algorithm used to perform neutrino propagation. However, the main scope of the article is different. Our objective is to extend the phase-space method, 
originally proposed in Ref. \cite{Lac22} for two neutrino beams to treat the problem of many beams, with varying numbers of neutrinos per beam ranging from very few to any large number of particles. The approach is validated using available existing exact calculations, indicating very 
good predictive power while allowing the description of a very large number of neutrinos. 
After presenting the Phase-Space Approach (PSA) and various new aspects introduced in this work, we show several examples of applications changing the number of beams and/or the number of neutrinos per beam.

\section{Hamiltonian simulation of neutrino oscillations } 

We consider an ensemble of neutrinos and assume that each neutrino can oscillate between different flavors. 
Specifically, we are interested in neutrinos emitted by stellar objects like neutron stars and, following previous work \cite{Peh11}, we 
assume that particles can be grouped into subsets having similar properties, i.e., similar initial states, momentum, and energy. 
Each group of neutrinos forms a "neutrino beam" where, within a beam, neutrinos are permutation invariant. This simplified approximation is particularly suitable to benchmark the phase-space approach that we discuss here. Indeed, as will be illustrated below, it was widely used 
previously to focus on the effect of neutrino-neutrino interaction, like entanglement. In addition, exact solution can be numerically performed for not too large beam/neutrino numbers, which will be crucial here to benchmark our approach. Although some critical discussion on this Hamiltonian and on the beam approach 
have been recently made \cite{Sha23,Joh23,Koh24}, and more general Hamiltonian have been recently proposed \cite{Cir24}, testing the PSA approach to the "neutrino beam" Hamiltonian is an important milestone for 
this approach. It is also a crucial test for pushing the limit of classical computer compared to quantum computer capabilities where this Hamiltonian can be can regarded as a pilot application.
Noteworthy, and although it is beyond the scope of the article, the PSA approach can a priory accommodate with more general or alternative many-body formulations. 

In the neutrino beam approximation, the total system is formed by $\alpha = 1, \cdots, n_B$ beams, and each beam $\alpha$ contains a number $N_{\alpha}$ of neutrinos.   
In the present work, we assume the so-called two-flavor approximation. We denote by 
$\{ | 0,i, \alpha  \rangle; | 1,i, \alpha \rangle\}$  the two eigenstates of the mass (mass basis) associated with the neutrino $i$ in the beam $\alpha$. The notation $\{ | f_{0,i, \alpha}  \rangle; | f_{1,i, \alpha} \rangle\}$ is introduced for the two states in the flavor basis.  These states are associated respectively with the two sets of creation operators $\{ a^\dagger_{0,i}(\alpha) , a^\dagger_{1,i}(\alpha) \}$ [mass basis] and $\{ f^\dagger_{0,i}(\alpha) , f^\dagger_{1,i}(\alpha) \}$  [flavor basis]. Within a beam, the Bogolyubov transformation from one basis to another is described by two angles $(\theta^f_\alpha, \phi^f_\alpha)$ and reads:
\begin{eqnarray}
\left\{ 
\begin{array}{l}
 f^\dagger_{0,i} (\alpha)      =  \cos \left( \frac{\theta^f_{\alpha}}{2}\right)  a^\dagger_{0,i} (\alpha)+ \sin \left( \frac{\theta^f_{\alpha}}{2}\right) e^{i \phi^f_{\alpha}} a^\dagger_{1,i} (\alpha)  \\
 \\
 f^\dagger_{1,i} (\alpha)  = - \sin \left( \frac{\theta^f_{\alpha}}{2}\right) e^{ - i \phi_{f,\alpha}}  a^\dagger_{0,i} (\alpha) + \cos \left( \frac{\phi^{f}_{\alpha} }{2}\right)  a^\dagger_{1,i} (\alpha)         
\end{array}
\right. 
. \label{eq:bogo}
\end{eqnarray}
Note that here, we use a general case with 2 angles defining 
the transformation between the mass and flavor basis, but in examples below, we
often assume $\phi^{f}_{\alpha} = 0$ and only one angle is used.

Within the two-flavor approximations, the neutrino oscillation problem
can be mapped onto a set of interacting two-level systems that can be identified as particles with  spins $1/2$ or qubits. 
Then, the Hamiltonian can be written as a function of the associated Pauli matrices acting on the two levels. A compact form 
of the Hamiltonian is given by:
\begin{equation}
    H = \sum_{\alpha=1}^{n_B} \omega_\alpha \vec{b} \cdot \vec{J}_{\alpha} + \sum_{\alpha \neq \beta}^{n_B} G_{\alpha \beta}(t) \vec{J}_{\alpha} \cdot \vec{J}_{\beta}\,,
    \label{eq:Hamiltonian}
\end{equation} 
where $\vec{J}_{\alpha} $ denotes the total spin vector of the  beam $\alpha$, defined as:
\begin{equation}
    \vec{J}_{\alpha} = (
        J_{\alpha}^x, J_{\alpha}^y, J_{\alpha}^z) \equiv \frac{1}{2}\sum_{i=1}^{N_{\alpha}} \vec{\sigma}_{i,\alpha}. \nonumber 
\end{equation}
Here $\vec{\sigma}_{i,\alpha} = (\sigma_x^{i, \alpha} , \sigma_y^{i, \alpha} ,\sigma_z^{i, \alpha} )$ denote the three Pauli matrices 
associated to the neutrino $i$ in the beam $\alpha$. Below, we will solve the problem on the mass basis, where these Pauli matrices are linked 
to the single-particle mass states according to:
\begin{eqnarray}
\left\{ 
\begin{array}{l}
\sigma^{i, \alpha}_x =   ~| 1, i , \alpha \rangle \langle  0, i , \alpha |  + ~ | 0, i , \alpha \rangle \langle  1, i , \alpha |  \\
\sigma^{i, \alpha}_y =  i  | 1, i , \alpha \rangle \langle  0, i , \alpha |  - i  | 0, i , \alpha \rangle \langle  1, i , \alpha |  \\
\sigma^{i, \alpha}_z =   ~| 0, i , \alpha \rangle \langle  0, i , \alpha |  -  ~| 1, i , \alpha \rangle \langle  1, i , \alpha |  
\end{array}
\right. . \label{eq:paulim}
\end{eqnarray} 
The one-body term in the Hamiltonian (\ref{eq:Hamiltonian}) depends on 
the vacuum oscillation frequency $\omega_\alpha=\Delta m^2_{01}/(2 E_{\alpha})$, which is expressed in terms of the difference in squared masses of neutrinos
$\Delta m^2_{01}$ and of the beam energy $E_\alpha$. $\Vec{b}$ in the mass basis identifies with $(0,0,-1)$. 

The second term in (\ref{eq:Hamiltonian}) describes the two-body interaction between neutrinos after emission from a compact stellar object. The two-body strength interaction is given by 
\begin{eqnarray}
G_{\alpha \beta}(t) &=&  \frac{\mu}{N} \left[ 1 - \cos(\theta_{\alpha \beta}(t)) \right ],  \label{eq:coupling_matrix}
\end{eqnarray}
where the coupling constant $\mu=\sqrt{2}G_F\rho_\nu$ is proportional to Fermi's constant $G_F$, and to the neutrino number density $\rho_\nu$. $N= \sum_\alpha N_\alpha$ is the total number 
of neutrinos. 
$\theta_{\alpha \beta}(t)$ is the relative angle between two neutrino beams $(\alpha, \beta)$ that tends to zero as neutrinos escape from 
their emitter, also leading to $G_{\alpha \beta}(t) \longrightarrow 0$  at large distance. 
Notably, MSW effect that describes the interaction between neutrinos and matter before emission is neglected here 
\cite{Wol78,Mik85}. In the first part of this work, we will assume that the set of angles $\{ \theta_{\alpha \beta} \}$ does not depend on time, corresponding to a time-independent 
Hamiltonian. This hypothesis will be relaxed in section \ref{sec:timedependent}.  

One of the great advantages of treating the neutrino problem as a set of coupled spins is that the permutation invariance within each beam
is automatically accounted for using the standard 
$\{ | J, M \rangle_\alpha \}_{M=-J, \cdots, +J}$ basis associated to the total spin $|\vec J|^2$ and $J_z$ component 
of the beam $\alpha$ (with $J=N_\alpha/2$).  This significantly reduces the numerical effort to perform 
a full configuration interaction (FCI) exact treatment of interacting neutrinos. In recent years, several FCI applications 
have been made, giving insight into the effect of beam interaction on entanglement and thermalization 
\cite{Mar21,Xio22,Ill22,Lac22, Cer22, Rog22a, Mar23a,Mar23b,Bha23}. This brute-force numerical solution, which takes advantage of the problem's symmetry, 
still becomes rapidly prohibitive as the number of beams increases and/or as the number of neutrinos in each beam 
also increases. 
Applications of FCI have been made assuming several beams but with only one neutrino per beam (see, for instance, \cite{Mar23a,Mar23b,Bha23}). When more neutrinos are considered within a beam, as far as we know, only the case of two and, more recently, three beams \cite{Rog22a} have been studied.     
When the number of neutrinos per beam and/or the number of beams increases, quantum computers can be the only alternative to perform exact 
simulations. Some illustration of neutrino evolution obtained on the IBM emulator of quantum computers will be shown in the present work. 
Approximate methods have also been explored, offering better numerical scaling with the number of particles such as Bethe ansatz \cite{Peh11,Bir18,Pat19}  or tensor network \cite{Rog21,Cer22}.    
These techniques remain rather involved when increasing the number of beams/neutrinos. In Ref. \cite{Lac22}, an alternative 
approach, called Phase-Space Approach (PSA), has been proposed, which keeps the simplicity of mean-field theory, successfully reproduces 
the FCI calculation effect of correlations and entanglement for the case of two interacting beams. One of the objective of this work is to further assess its 
predictive power in more general situations. 

\subsection{Phase-Space Approach to neutrino oscillations}

The PSA technique was originally proposed in the nuclear physics model to circumvent the failure of mean-field to describe quantum fluctuations beyond mean-field \cite{Ayi08,Lac12} (see also the review \cite{Lac14}).  Several successful applications have been made so far in different fields of physics \cite{Lac12, Lac13, Yil14, Lac14b, Lac16, Ulg19}.  
There are two important ingredients in this approach: (i) The quantum fluctuations of the initial system can be mapped into a statistical 
initial sampling problem leading to a set of initial conditions that, on average, reproduces the exact quantum problem,  and (ii) Each set of the initial 
sampling can be evolved independently from the others using simple equations of motion (EoM) that are supposed to identify with the time mean-field 
EoM. This technique is efficient, provided the interferences between different mean-field trajectories are not too strong \cite{Reg18} (see Refs.  
\cite{Czu20}  for extension of PSA allowing to correct for the missing interferences). 
A recurrent observation is that, even when these interferences play a non-negligible role, the approach still provides the correct early stage entropy growth and average asymptotic behavior,  
and might be useful for understanding thermalization or decoherence processes.  Here, we first recall the mean-field EoM that will be used to perform 
the evolution after sampling and then focus on the sampling itself. 

\subsubsection{Mean-Field equations of motion}

The exact evolution of the different spins $\{ \vec J_\alpha \}_{\alpha=1,n_B}$ expectations values are given by:
\begin{eqnarray}
i \hbar \frac{\mathrm{d} \langle \vec J_{\alpha} \rangle }{\mathrm{d}t} &=& \langle [ \vec J_{\alpha} , H ] \rangle, 
\end{eqnarray}  
that gives: 
\begin{eqnarray}
\displaystyle \frac{\mathrm{d}}{\mathrm{d}t} \langle \vec J_\alpha \rangle =  \omega_\alpha  \vec b \wedge \langle \vec J_\alpha \rangle  
+ \sum_{\beta \neq \alpha} G_{\alpha \beta}(t) \langle \vec J_\beta \wedge 
\vec J_\alpha \rangle.  \label{eq:eomone} 
\end{eqnarray} 
These equations show that the second moments of the spins appear on the left-hand side. The exact solution to the problem requires completing the evolutions of the average spin by the evolutions of their 
second-order moments. These moments are themselves 
coupled to the third moment, and so on and so forth.  Solving such coupled equations becomes rapidly prohibitively expensive, and some truncation schemes are generally necessary. The mean-field theory is the simplest approximation and can be obtained by neglecting quantum fluctuations altogether, i.e. by assuming $\langle J^{x,y,z}_\alpha J^{x,y,z}_{\beta}\rangle \simeq \langle J^{x,y,z}_\alpha \rangle \langle  J^{x,y,z}_{\beta}  \rangle$.  

Defining the polarization vector components for each beam 
as $\vec P_\alpha = 2  \langle \vec J_\alpha \rangle/N_\alpha = (P^\alpha_x,  P^\alpha_y, P^\alpha_z)$, the mean-field EoM can be written as
\begin{eqnarray}
\displaystyle \frac{\mathrm{d}}{\mathrm{d}t} \vec P_\alpha =  \omega_\alpha  \vec b \wedge \vec P_\alpha  
+ \frac{1}{2}\sum_{\beta \neq \alpha}^{n_B} N_\beta G_{\alpha \beta}(t) \vec P_\beta   \wedge \vec P_\alpha.  \label{eq:eom-mf}
\end{eqnarray} 
The mean-field approximation is simple to solve numerically compared to the original problem since, for each beam, it only requires following 
the corresponding three polarization vector components, independently from the number of neutrinos in the beam.  However, neglecting quantum fluctuations 
beyond the mean field is a drastic approximation that prevents the proper description of true many-body effects or entanglement between particles. Such entanglement happens in the physics of neutrino oscillations when neutrinos interact with each other (see illustrations in \cite{Mar21,Lac22} for the case of two beams).  


\subsubsection{Mapping initial quantum fluctuations to statistical fluctuations}

A method that turns out to be rather efficient in treating quantum fluctuations while keeping 
the simplicity of mean-field EoM is to replace the above deterministic method with a stochastic 
method leading to a set of mean-field trajectories to be considered \cite{Lac14}, where the randomness stems 
from the sampling of the initial conditions.   A key feature of the PSA approach is that quantum mean expectation values and 
their quantum variances identify with the mean and variance obtained by performing the classical average over 
the initial conditions. Explicitly, we consider a set of observables  $\{ \hat A_m \}$ and an initial state described by a density $D(0)$. 
These observables are usually one-body observables when many-body problems are considered, as is the case here. 
The quantum mean and variance of these operators at initial time are given by:
\begin{eqnarray}
\langle \hat A_m \rangle &=& {\rm Tr}(\hat A_m D(0)), \nonumber \\
\sigma^2_{A_m} &=& {\rm Tr}\left(\hat A^2_m D(0)\right) -   {\rm Tr}\left(\hat A_m D(0)\right)^2. \nonumber
\end{eqnarray}
When replacing the problem by a statistical ensemble, for each operator, a set of complex numbers denoted by 
$\{ A^{(\lambda)}_m \}_{\lambda =1, \cdots, N_{\rm evt}}$ is randomly generated. Here, $\lambda =1, \cdots, N_{\rm evt}$ labels the events, and the total number of events is $N_{\rm evt}$. A statistical average over the sampling then replaces the quantum expectation values of observables:
\begin{eqnarray}
\overline{A^{(\lambda)}_m} &=& \frac{1}{N_{\rm evt}} \sum_{\lambda=1}^{N_{\rm evt}} A^{(\lambda)}_m, ~~
\Sigma^2_{A_m} =
\overline{A^{(\lambda)}_m A^{(\lambda)}_m} -  \overline{ A^{(\lambda)}_m} ^2 \label{eq:devtat} 
\end{eqnarray}   
As pointed out in \cite{Yil14}, the sampling strategy is not unique, but it is constrained to reproduce the first and second moments of the selected 
observables, i.e. 
\begin{eqnarray}
\overline{A^{(\lambda)}_m} &=&\langle \hat A_m \rangle , ~~ \Sigma^2_{A_m} = \sigma^2_{A_m}. \label{eq:meanvar}
\end{eqnarray} 
For the neutrino problem, the natural choice for the observables is the different total spin components for the different beams, leading to a set
of initial conditions $\vec J^{(\lambda)}_{\alpha}$, or equivalently a set of initial 
values for the polarization vector components $\left\{ \left({ P}^{\alpha(\lambda)}_x(0) , {P}^{\alpha(\lambda)}_y(0), {P}^{\alpha(\lambda)}_z(0) \right) \right\}_{\alpha = 1,\cdots,n_B}$. 
Each initial sample is then evolved according to its own mean-field evolution given by Eq. (\ref{eq:eom-mf}). Then, any mean values or fluctuations 
are calculated through a classical average of observables along the different sampled trajectories.  There are several great advantages of the PSA technique:
\begin{itemize}
  \item First, the stochastic process only arises from the initial conditions, considerably limiting the number of events sampled to achieve statistical accuracy for observable calculations. 
  This is at variance with most other quantum stochastic methods, such as quantum Monte Carlo or quantum state diffusion \cite{Cep95,Fou01,Car15}.   
  
  \item Ultimately, only mean-field evolution is needed and this scales linearly with the number of observable retained. This has to be compared with the exponential
  increase of the number of degrees of freedom to follow when treating exactly a many-body problem. In the neutrino case with two flavors, each event requires to solve 
  $3n_B$ non-linear coupled equation, independently of the number of neutrinos per beam. Noteworthy, going to the three-flavor case would lead to $8n_B$ coupled equations.        
  \item Finally, since the trajectories are independent of each other, calculations can be straightforwardly parallelized by sending different trajectories on different 
  CPUs.     
\end{itemize} 
Despite its apparent simplicity, the PSA approach is generally able not only to describe quite accurately the propagation of fluctuations where mean-field alone fails but also to incorporate the effect of fluctuations on the one-body evolution.  It was shown in Ref. \cite{Lac22}, for instance, to be able to describe the one- and 
two-neutrino entanglement entropies. As shown in Ref. \cite{Lac16, Czu20}, this success stems from the fact that the PSA approach is 
equivalent to solving a non-truncated 
BBGKY hierarchy \cite{Bog46,Bor46,Kir46,Bon16}, where the difference with the exact BBGKY only arises at the level of two-body observables due to 
the neglect of the Pauli exclusion principle in some terms. Additionally, it is interesting to mention that, in the context of open quantum systems, the fluctuations and dissipation terms appearing in quantum Brownian motion are obtained from the absence of knowledge of the initial condition for the environment that is treated by a sampling of the environment's initial conditions that propagate in time (see for instance the discussion in chapter 3 of Ref. \cite{Bre02}).  A similar idea is proposed in PSA with the great difference that we consider a closed finite many-body system here.          

In this section, we highlighted the philosophy and strategy behind the PSA method. Below, we focus on the practical aspects
of sampling, and several applications are given, proving its predictive power in the context of neutrino physics. Two major improvements are made compared to our previous applications: (i) We explored several methods to perform the initial sampling, in particular, to be able to apply the PSA method from the extreme case of one particle per beam to many; (ii) we extend the method to be able to treat an arbitrary number of beams $n_B$. In Ref. \cite{Lac22}, only the case $n_B=2$ with at maximum $50$ neutrinos per beam was considered

\subsubsection{Different methods for sampling initial conditions}

Most applications using simplified Hamiltonian given by Eq. (\ref{eq:Hamiltonian}) assume that different beams are initially uncorrelated and that each beam's initial states identify with a coherent state. We focus here on this case. The initial state is then given by: 
\begin{eqnarray}
    | \Psi(0) \rangle &=& \bigotimes_{\alpha=1}^{n_B} | \Omega_\alpha \rangle, 
\end{eqnarray}
where $\Omega_\alpha = (\theta_\alpha, \phi_\alpha)$, with $\theta_\alpha \in [0,\pi]$ and $\phi_\alpha \in [0, 2 \pi]$, defines an orientation in the Bloch sphere. 
Starting  from the mass basis, a SU(2) coherent state $| \Omega_\alpha \rangle$ corresponds to a Slater determinant written as:
\begin{eqnarray}
|\Omega_\alpha \rangle = \prod_{i=1}^{N_\alpha} c^\dagger_{i, \alpha}  |- \rangle. \label{eq:coherent-alpha}
\end{eqnarray}
Here $|-\rangle$ stands for the Fock space vacuum. Note that in the PSA approach, a more general initial state can be considered like systems at finite temperature or initially correlated systems (see, for instance, Refs. \cite{Ayi08,Lac14,Yil14}). We, however, restrict to this specific case here because this was the assumption made in most of the recent studies aiming at solving exactly the neutrino beam problems.

The creation operators in Eq. (\ref{eq:coherent-alpha}) obey 
similar transformations as the one given in Eq. (\ref{eq:bogo}) and creates a set of single-particle states denoted by $|\varphi_{i,\alpha}\rangle$. Eventually, these states identify with one of the flavor states if $(\theta_\alpha, \phi_\alpha) = (\theta^f_\alpha, \phi^f_\alpha)$ [$0$ flavor state] or $(\theta_\alpha, \phi_\alpha) = (\pi -\theta^f_\alpha, \phi^f_\alpha)$ [$1$ flavor state] according to the convention of
Eq. (\ref{eq:Hamiltonian}).    

Because the initial state is a tensor product of different beam states, sampling different polarization components 
can be made independently for each beam. Three different methods have been employed in the present work, the last one
being new. These methods are called hereafter Gaussian, Husimi, and bi-valued samplings and are detailed below.  

\noindent {\bf Gaussian sampling:} The Gaussian sampling technique was the original sampling method implemented in PSA. 
It calculates the quantum mean and variance of the observables of interest. 
Then, the sampling of the initial values of the observables is made assuming a multi-dimensional Gaussian probability distribution 
where the mean values and variances both identify with the quantum ones, as imposed by Eq. (\ref{eq:meanvar}). For a given coherent state $| \Omega_\alpha \rangle$, it is first convenient 
to introduce SU(2) generators $\vec {\cal J}_\alpha = (\vec {\cal J}^\alpha_x, \vec {\cal J}^\alpha_y, \vec {\cal J}^\alpha_z)$ in the rotated frame attached to the state.  Note that here we refer to the original frame, the one associated to the creation operators $\{ a^\dagger_{0,i}(\alpha),a^\dagger_{1,i}(\alpha)\}$, i.e. the mass frame.   
The two sets of operators, i.e. the one in the original and rotate frames are linked through: 
\begin{widetext}
\begin{eqnarray}
\left\{ 
\begin{array}{l}
{\cal J}^\alpha_{x} = +\left[ c^2_\alpha -s^2_\alpha \cos(2 \phi_\alpha) \right] J^\alpha_x  - s^2_\alpha \sin(2 \phi_\alpha) J^\alpha_y   - \sin(\theta_\alpha) \cos(\phi_\alpha)J^\alpha_z,  \\
\\
{\cal J}^\alpha_{y} = -s^2_\alpha   \sin(2 \phi_\alpha) J^\alpha_x  + \left[ c^2_\alpha +s^2_\alpha \cos(2 \phi_\alpha) \right] J^\alpha_y  - \sin(\theta_\alpha) \sin(\phi_\alpha) J^\alpha_z,  \\
\\
{\cal J}^\alpha_{z}= \sin(\theta_\alpha) \cos(\phi_\alpha) J^\alpha_x + \sin(\theta_\alpha) \sin(\phi_\alpha) J^\alpha_y   + \cos(\theta_\alpha) J^\alpha _z. 
\end{array}
\right. \label{eq:masstoflavor}
\end{eqnarray}
with the compact notations: $c_\alpha = \cos(\theta_\alpha/2)$ and
 $s_\alpha = \sin(\theta_\alpha/2)$. We also have the inverse transformation:
\begin{eqnarray}
\left\{ 
\begin{array}{l}
J^{\alpha} _x = +\left[ c^2_\alpha -s^2_\alpha \cos(2 \phi_\alpha) \right] {\cal J}^{\alpha}_{x} - s^2_\alpha \sin(2 \phi_\alpha) {\cal J}^\alpha_{y} + \sin(\theta_\alpha) \cos(\phi_\alpha) {\cal J}^{\alpha}_{z},  \\
\\
J^{\alpha} _y  = -s^2_\alpha   \sin(2 \phi_\alpha) {\cal J}^{\alpha}_{x} + \left[ c^2_\alpha +s^2_\alpha \cos(2 \phi_\alpha) \right] {\cal J}^{\alpha}_{y} + \sin(\theta_\alpha) \sin(\phi_\alpha){\cal J}^{\alpha}_{z},   \\
\\
J^{\alpha} _z  = -  \sin(\theta_\alpha) \cos(\phi_\alpha) {\cal J}^{\alpha}_{x}- \sin(\theta_\alpha) \sin(\phi_\alpha) {\cal J}^{\alpha}_{y} + \cos(\theta_\alpha) {\cal J}^{\alpha}_{z}.  
\end{array}
\right. \label{eq:flavortomass}
\end{eqnarray}
 \end{widetext}
The state $| \Omega_\alpha \rangle$ corresponds to the eigenstate $| {\cal J}_\alpha, - {\cal J}_\alpha \rangle$ in the rotated frame. In this frame, it has simple mean and fluctuations 
properties:
\begin{eqnarray}
\left\{
\begin{array}{l}
\displaystyle  \langle {\cal J}^\alpha_x \rangle = \langle {\cal J}^\alpha_y \rangle = 0, ~~ \langle {\cal J}^\alpha_z \rangle = - {\cal J}_\alpha = -\frac{N_\alpha}{2} ,  \\
\\
\displaystyle 
\sigma^2_{{\cal J}^\alpha_x} = \sigma^2_{{\cal J}^\alpha_y} = \frac{N^2_\alpha}{4}, ~~ \sigma^2_{{\cal J}^\alpha_z} = 0 . 
\end{array}
\right.  \label{eq:gaussampling}
\end{eqnarray} 
The sampling is then performed following the scheme:
\begin{itemize}
  \item A set of initial values $({\cal J}^{\alpha(\lambda)}_{x}, {\cal J}^{\alpha(\lambda)}_{y}, {\cal J}^{\alpha(\lambda)}_{z})$ is generated assuming 
  that the two first components follow a Gaussian probability with mean zero and variance equal to $N^2_{\alpha}/4$, while ${\cal J}^{\alpha(\lambda)}_{z}$
  is a non-fluctuating quantity equal to $-N_\alpha/2$.
  \item These components are then transformed back to the original frame using Eq. (\ref{eq:flavortomass}), leading to a set of events $({ J}^{\alpha(\lambda)}_{x}, {J}^{\alpha(\lambda)}_{y}, {J}^{\alpha(\lambda)}_{z})$. 
  \item The initial values for the polarization vector are then obtained simply using $\vec P^{(\lambda)}_\alpha = 2 \vec J_{\alpha}^{(\lambda)}/N_\alpha$. 
  \item Each initial polarization vector is then evolved according to the mean-field equation (\ref{eq:eom-mf}). 
\end{itemize} 

\noindent{\bf Husimi sampling:} This sampling was originally proposed in Ref. \cite{Yil14} and is based on the Husimi probability distribution (also called Q-probability). 
In particular, this approach does not require going back and forth from the mass to the rotated frame. The Husimi distribution associated with a coherent state is given by  
\begin{eqnarray}
Q_\alpha (\theta, \phi) 
&=& \left[ \frac{1 +  \cos\theta  \cos \theta_\alpha +  \sin \theta \sin\theta_\alpha \cos(\phi -\phi_\alpha)}{2} \right]^{N_\alpha} .\nonumber 
\end{eqnarray} 
In practice, a set of values of $(\theta^{(\lambda)},\phi^{(\lambda)})$ values can be obtained using a 
Metropolis sampling with this probability distribution.   One subtle aspect of this approach is that 
the expectation value of an observable $O$ acting on the beam $\alpha$ is obtained through the average:
\begin{eqnarray}
\langle  O \rangle &=& \frac{N_\alpha + 1}{4\pi} \int_\Omega W_O (\Omega) Q_\alpha(\Omega) \mathrm{d}\Omega.  \label{eq:husimimean}
\end{eqnarray} 
where $W_O (\Omega) $ is the expectation value of the Weyl operator associated with the observable and denoted by $\hat W_O$, i.e., $W_O (\Omega)  = \langle \Omega |\hat W_O| \Omega \rangle$. For spin systems, we have specifically \cite{Gil76}
  \begin{eqnarray}
  W_{J^{\alpha}_{x,y,z}} &=&  \frac{N_\alpha+2}{N_\alpha} J^\alpha_{x,y,z} . \label{eq:weyl1} 
  \end{eqnarray} 
 In particular, using the Metropolis sampling technique and  expression (\ref{eq:weyl1}), the sampling of angles leads to three 
 random variables:
 \begin{eqnarray}
W^{(\lambda)}_{J^{\alpha}_{x}} &=&  \frac{N_\alpha+2}{2}   \sin(\theta^{(\lambda)}) \cos(\phi^{(\lambda)}), \nonumber \\
W^{(\lambda)}_{J^{\alpha}_{y}} &=&   \frac{N_\alpha+2}{2}  \sin(\theta^{(\lambda)}) \sin(\phi^{(\lambda)}),    \nonumber \\
W^{(\lambda)}_{J^{\alpha}_{z}} &=&  \frac{N_\alpha +2}{2} \cos(\theta^{(\lambda)}). \nonumber 
\end{eqnarray}     
The classical average over events of these quantities properly matches the expectation value of the $\vec J_\alpha $ components, but it does not reproduce the fluctuations. 
To be able to take advantage of the Husimi technique and still properly describe the first and second moments of the spin components, we follow 
the prescription of Ref. \cite{Yil14} and define the component $\vec J^{(\lambda)}_\alpha$ according to:
\begin{eqnarray}
{ J}^{\alpha(\lambda)}_{x,y,z} &=& W^{(\lambda)}_{J^{\alpha}_{x, y, z}} + \sqrt{\frac{\sigma^2_{J_{x,y,z}}}{\Sigma^2_{x,y,z}}} \delta W^{(\lambda)}_{x,y,z},
\end{eqnarray}   
where we use the short-hand notations:
\begin{eqnarray}
\delta W^{(\lambda)}_{x,y,z} &\equiv&  W^{(\lambda)}_{J^{\alpha}_{x, y, z}} - \langle J^{\alpha}_{x, y, z} \rangle \nonumber 
\end{eqnarray}
and
\begin{eqnarray}
\Sigma^2_{x,y,z} &\equiv& \overline{\delta W^{(\lambda)}_{x,y,z} \delta W^{(\lambda)}_{x,y,z} }. \nonumber
\end{eqnarray} 
This procedure consists in renormalizing the fluctuations to match the quantum fluctuations while keeping the first moment unchanged. 
Again, each initial set of $({ J}^{\alpha(\lambda)}_{x}, { J}^{\alpha(\lambda)}_{y}, { J}^{\alpha(\lambda)}_{z})$ is then used to obtain initial values of the polarization followed by mean-field EoM. 

\noindent{\bf Measurement-inspired sampling or "bi-valued sampling":} Besides the approximation made by mapping a quantum problem into a statistical problem, 
one potential shortcoming of the two previous sampling techniques is that the sampling only ensures that the two first moments of the 
distribution are reproduced.  The Husimi is expected to be less constraining since it does not presuppose that the probability distribution is Gaussian but 
is close to the Husimi quasi-probability. In the present work, we are eventually interested in the extreme limit of one particle per beam for which the Gaussian limit dramatically
breaks down. Indeed,  considering this limit, the spin components $\vec {\cal J}_\alpha = (\vec {\cal J}^\alpha_x, \vec {\cal J}^\alpha_y, \vec {\cal J}^\alpha_z)$  
in the rotated frame, becomes simply proportional to the Pauli matrices $(X_\alpha, Y_\alpha, Z_\alpha)$ in this frame. In particular, Pauli matrices verify:
\begin{eqnarray}
O^{2k}_\alpha &=& I_\alpha, ~~ O^{2k+1}_\alpha = O_\alpha
\end{eqnarray}   
where $O \subset \{ X, Y, Z\}$, and $I_\alpha$ is the identity operator. These properties invalidate the Gaussian approximation as can be seen by computing different moments. For instance, the fourth-centered moment of any $O_\alpha$ is not $3$ times the square of the second-centered moment as expected for a Gaussian. A distribution properly describing the different moments can be inspired by measurement theory and quantum state 
tomography in quantum computing \cite{Nie02,Aar18,Hua20,Rui24}.    

Let us consider that we transform the spin of a single neutrino into a qubit using an SU(2) mapping as in Refs. \cite{Hal21,Ami23}
with the convention that $|s,s_s \rangle = \{ | \frac{1}{2}, -\frac{1}{2} \rangle, | \frac{1}{2}, \frac{1}{2} \rangle \}$ becomes $\{|0 \rangle , |1\rangle \}$. 
For a single neutrino per beam, the initial state in the rotated frame is $|0 \rangle$. Since the qubit register is already in the $Z$-basis by convention, one can 
measure it repeatedly. In this simple case, one will always get a $0$. One can generate a single-valued probability denoted 
by $P_Z(r) = \delta(r)$ by measurement.  One can do the same for the $X$ and $Y$ operators, with the difference that one should go from the 
$Z$ basis to the $X$ or $Y$ basis by performing a unitary transformation before the measurement. In both cases, provided that the state to measure 
is $|0 \rangle$, one would get a random set of values in $\{ 0, 1 \}$, with probabilities to measure zero or one equal to $1/2$. 
Averaging over the measurements will properly match all moments $\langle X^k\rangle$ or $\langle Y^k\rangle$ if the measurement is made respectively 
in the $X$- or $Y$-basis. 
Quantum 
measurement theory leads to the simple conclusion that the generation of a set of random numbers $r^{(\lambda)}$  for the $X$ and $Y$ components can be made from the bi-valued  probability 
\begin{eqnarray}
P_{X/Y}(r) &=& \frac{1}{2} \left( \delta(r) + \delta(r-1) \right) \label{eq:bi-valued}
\end{eqnarray}
Note that the possibility of using a bi-valued distribution has already been discussed in Ref. \cite{Ulg19} using second-quantization properties. 
Using measurement arguments gives a simple prescription for sampling initial polarization components for the case of a single neutrino per beam. In the rotated frame, we can directly assume 
that 
\begin{eqnarray}
({\cal J}^{\alpha(\lambda)}_{x}, {\cal J}^{\alpha(\lambda)}_{y}, {\cal J}^{\alpha(\lambda)}_{z})
 &\equiv& -\frac{1}{2}\left({x}_\alpha^{(\lambda)}, y_\alpha^{(\lambda)}, 1 \right) \nonumber
\end{eqnarray}
where $(x^{(\lambda)}_\alpha,y^{(\lambda)}_\alpha)$ are two random variables taking the values $(1,-1)$ with probabilities $1/2$ for each values. This sampling 
can be extended to the case of $N_\alpha$ qubits per beam simply by generating $2N_\alpha$ random numbers 
$\{ (x^{(\lambda)}_{i ,\alpha}
    , y^{(\lambda)}_{i,\alpha}) \}_{i=1,\cdots, N_\alpha} $ each having bi-valued probabilities, such that  we have more generally
\begin{eqnarray}
({\cal J}^{\alpha(\lambda)}_{x}, {\cal J}^{\alpha(\lambda)}_{y}, {\cal J}^{\alpha(\lambda)}_{z})
 &\equiv& -\frac{1}{2} \left\{ \sum_{i=1}^{N_\alpha} ({x}_\alpha^{(\lambda)}, y_\alpha^{(\lambda)}, 1 ) \right\} .
\end{eqnarray}
Again, one can transform the total spin in the mass frame and perform the evolution with the mean-field EoM. Note that, 
due to the central limit theorem, since both  ${\cal J}^{\alpha(\lambda)}_{x}$ and ${\cal J}^{\alpha(\lambda)}_{y}$ are sums
of random variables, their probability distribution will tend to a Gaussian distribution as $N_\alpha$ increases.     

\begin{figure}[htbp] 
  \includegraphics[scale=0.38]{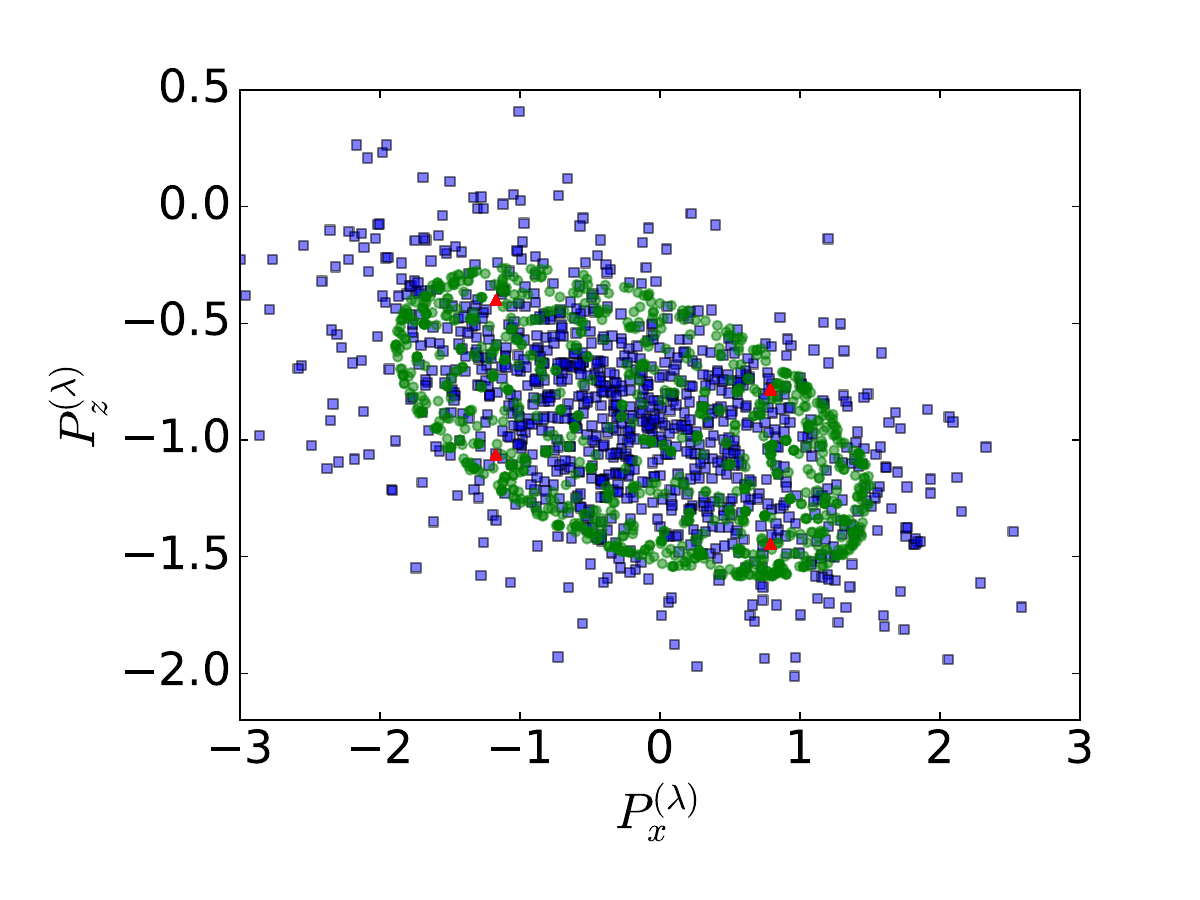} 
    \caption{Illustration of the $(P^{(\lambda)}_x, P^{(\lambda)}_z)$ components in the mass basis obtained for one beam with only one neutrino initialized in the coherent state $| \theta = \pi/8, \phi = \pi/3 \rangle$ using the Gaussian sampling (blue squares), Husimi sampling (green circles), and bi-valued sampling (red triangles). For each sampling method, $1000$ events are presented. The appearance of only four red squares in Fig. 1 might appear mysterious at first sight. In this sampling technique, all generated events falls down in one of these four locations in the graph. This stems from the very discretized nature of the bi-valued sampling. With this sampling, the proper mean values and widths of operators are obtained by the statistical average of the 4 points that are initially sampled with different weights.}
    \label{fig:sampling}
\end{figure}

As an illustrative example of the distribution obtained with the different sampling methods, we show in Fig.  \ref{fig:sampling} 
a comparison of a set of events generated with the three approaches for a single neutrino per beam case. As can be seen from the figure, 
even though all probabilities have the same mean values 
and fluctuations for the polarization components, the initial sampling
strongly depends on the method used.  

\section{Applications}

The present work has several objectives: (i) Firstly, we would like to validate the PSA approach for cases with $n_B >2$; (ii) Secondly, 
when exact solutions are doable, we would like to compare them with PSA, and further prove the predictive power of the approach, especially to reproduce 
many-body properties like entanglement. Provided that these two objectives are reached, the PSA is promoted to an extremely competitive
tool to simulate the neutrino beam evolution problem on a classical computer.

In recent years, several applications of the interacting beam problem have been made on quantum computers, assuming only one neutrino per beam \cite{Hal21,Ami23}. These applications are restricted to a rather low number
of beams ($n_B \lesssim 20$) \cite{Mar23a,Mar23b} on quantum computer emulators due to the numerical cost that    
 rapidly exceeds the capability of classical computers. In real quantum devices, with the extra constraint of device noises, most recent simulations 
 have been made up to $n_B \le 8$ over a rather short time \cite{Ami23}. For a larger number of particles per beam, a few applications have been 
 made for $n_B=2$ beams \cite{Mar21,Xio22}, and, as far as we know, only one application for $n_B=3$ was made in ref. \cite{Rog22a}. 
Since the PSA approach has already been validated for $n_B=2$ in Ref. \cite{Lac22}, we will concentrate on larger $n_B$ and 
use Refs \cite{Rog22a, Ami23} as benchmarks.  

\subsection{Interacting beam problem with one particle per beam}
\label{sec:1neutrino}
  
For the case of one particle per beam, we use the Hamiltonian proposed in Ref.  \cite{Ami23}. This hamiltonian 
can be written in the form (\ref{eq:Hamiltonian}), where the coupling matrix elements are given by:
\begin{eqnarray}
    G_{\alpha,\beta}(t)&=&\frac{G(t)}{n_B}\left(1-\cos\left[ \frac{|\alpha-\beta|}{(n_B-1)} \arccos(0.9)\right] \right),
    \label{eq:param}
\end{eqnarray}
where we directly identify the beam label $\alpha$ with the index on the neutrinos $i$. Note that here $n_B$ identifies with the total number of particles $N$ if we consider one particle per beam. 
Consistently with Ref. \cite{Ami23}, the flavor to mass 
conversion assumes $(\theta^f_\alpha, \phi^f_\alpha) = (2 \times 0.195, 0)$, where the factors $2$ stems from the convention in the Bogolyubov 
transformation (\ref{eq:bogo}). For instance, $\vec b =  ( \sin(\theta^f_\alpha), 0, -\cos(\theta^f_\alpha))$ with our convention.

 \begin{figure}[htbp] 
\includegraphics[width= \linewidth]{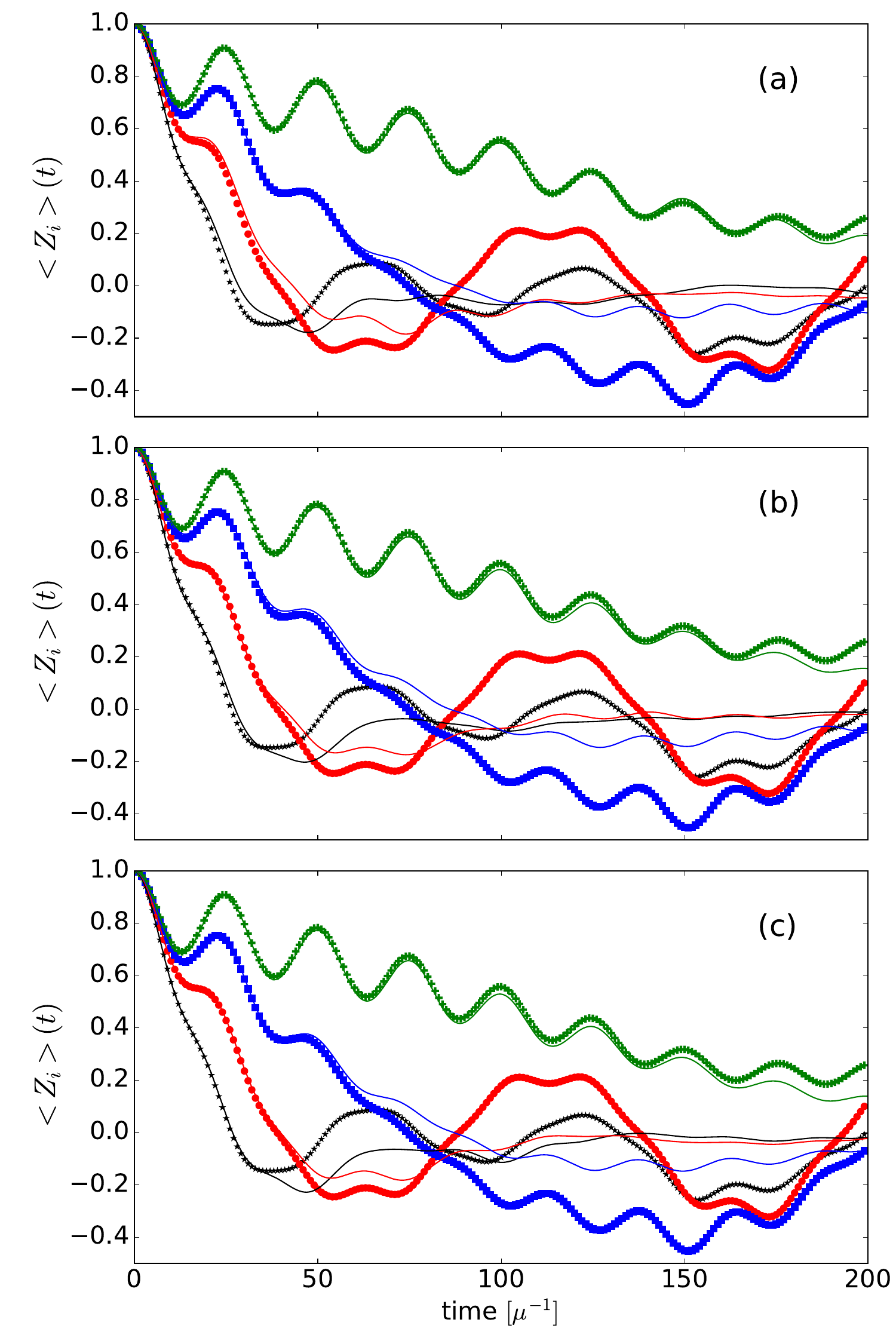} 
    \caption{Exact evolution of the quantity $\langle {\cal Z}_i \rangle$ components, estimated in the flavor basis, as a function of time (filled symbols) for the case of $n_B=8$ coupled neutrinos.
     Different symbols correspond to different neutrinos where half of the neutrinos are initialized in one of the flavor states, while the other half are initialized in the other. Exact evolutions (symbols) are compared to the phase-space approximation (solid lines) using Gaussian (a), Husimi (b), and bi-valued (c) sampling for the initial conditions. The reference calculations are obtained using the quantum computer algorithm performed with the qiskit emulator \cite{qiskit2024}. To simplify the figure we only show the neutrinos for which initially $\langle {\cal Z}_i \rangle=1$. Due to the symmetry of the problem, those having  $\langle {\cal Z}_i \rangle=-1$ are symmetric with respect to the $y=0$ axis.}    
     \label{fig:zevolution}
\end{figure}

\subsubsection{time-independent Hamiltonian}

The calculations of Ref. \cite{Ami23} uses a time-independent Hamiltonian with $\omega_\alpha = 2 \mu /n_B$ and  $\frac{G(t)}{n_B} = 2 \frac{\mu}{n_B}$ with 
the additional convention that $\mu = 1$ which is equivalent to write energies in $\mu$ units and time in $\mu^{-1}$ units  \footnote{Note that, there is also a factor $2$ in the coupling matrix elements and $\omega_\alpha$. This factor stems from the fact
that $\sum_{\alpha \neq \beta}$ is considered compared to Ref. \cite{Ami23} and the fact that the Hamiltonian is written in terms of the spins $\vec J_{\alpha}$, not the neutrinos Pauli matrices.  }. 

For the initial conditions, to compare with the work of Ref. \cite{Ami23}, we will consider an even number of beams/neutrinos where the first half of the neutrinos 
are in one flavor state ($|0^f_{\alpha}\rangle$ for $\alpha = 1, \cdots, N_\alpha/2$, i.e. $\theta_\alpha =\theta^f_\alpha$), and the other half is in the second flavor state 
($|1^f_{\alpha}\rangle$ for $\alpha = N_\alpha/2, \cdots, N_\alpha$, i.e. $\theta_\alpha =\pi - \theta^f_\alpha$). 

We have performed systematic tests of the PSA method with varying $n_B$ values. The mean-field EoM is solved using a Runge-Kutta 2 (RK2) algorithm with a
numerical time step $\Delta t =  0.05 ~[\mu^{-1}]$. For each PSA calculation, the evolution is made by averaging over $N_{\rm evt} = 10^4$ trajectories, leading to very small statistical deviation due to the finite number of events. The associated error bars are too small to be represented on figures. 
We also developed two codes solving the exact solutions, one on a classical computer and one on the qiskit quantum computer emulator \cite{qiskit2024}, the two codes providing the same results as the one published in \cite{Ami23}. These reference codes can emulate up to $n_B=20$ beams with one neutrino per beam on a desktop computer. 

Since in this first application, only one neutrino per beam is considered, we omit the label "$i$" in the Pauli matrices $\langle \sigma^{i,\alpha}_{x,y,z} \rangle$ and use the notations $(\sigma^{i,\alpha}_{x}, \sigma^{i,\alpha}_{y}, \sigma^{i,\alpha}_{z}) \equiv (X_\alpha, Y_\alpha, Z_\alpha)$. We also introduce in the flavor basis 
the equivalent matrices $ ({\cal X}_\alpha, {\cal Y}_\alpha, {\cal Z}_\alpha)$. One can single out the properties of one specific neutrino/beam $\alpha$ by focusing on its one-body density matrix $\rho^\alpha$ obtained by tracing the total density over other neutrinos.  The $2\times2$ density 
matrix has components in the mass basis given by:
\begin{eqnarray}
\left\{ 
\begin{array}{l}
\displaystyle \rho^\alpha_{00}(t) = \frac{1}{2} ( 1 + P^\alpha_z(t)) \\
\\
\displaystyle \rho^\alpha_{11}(t)=  \frac{1}{2} ( 1 - P^\alpha_z(t))  \\
\\
\displaystyle \rho^\alpha_{01}(t) = \frac{1}{2} (P^\alpha_x(t) - i P^\alpha_y(t)) 
\end{array}
\right. . \label{eq:densityone}
\end{eqnarray}
In the exact case, the polarization vector is obtained from the expectation of the three Pauli matrices.  In the PSA case, one computes these
quantities from the statistical average given in Eq. (\ref{eq:devtat}) over trajectories. For instance, we have $P^{\alpha}_x = \overline{P^{\alpha (\lambda)}_x}$.  

 \begin{figure}[htbp] 
\includegraphics[width=\linewidth]{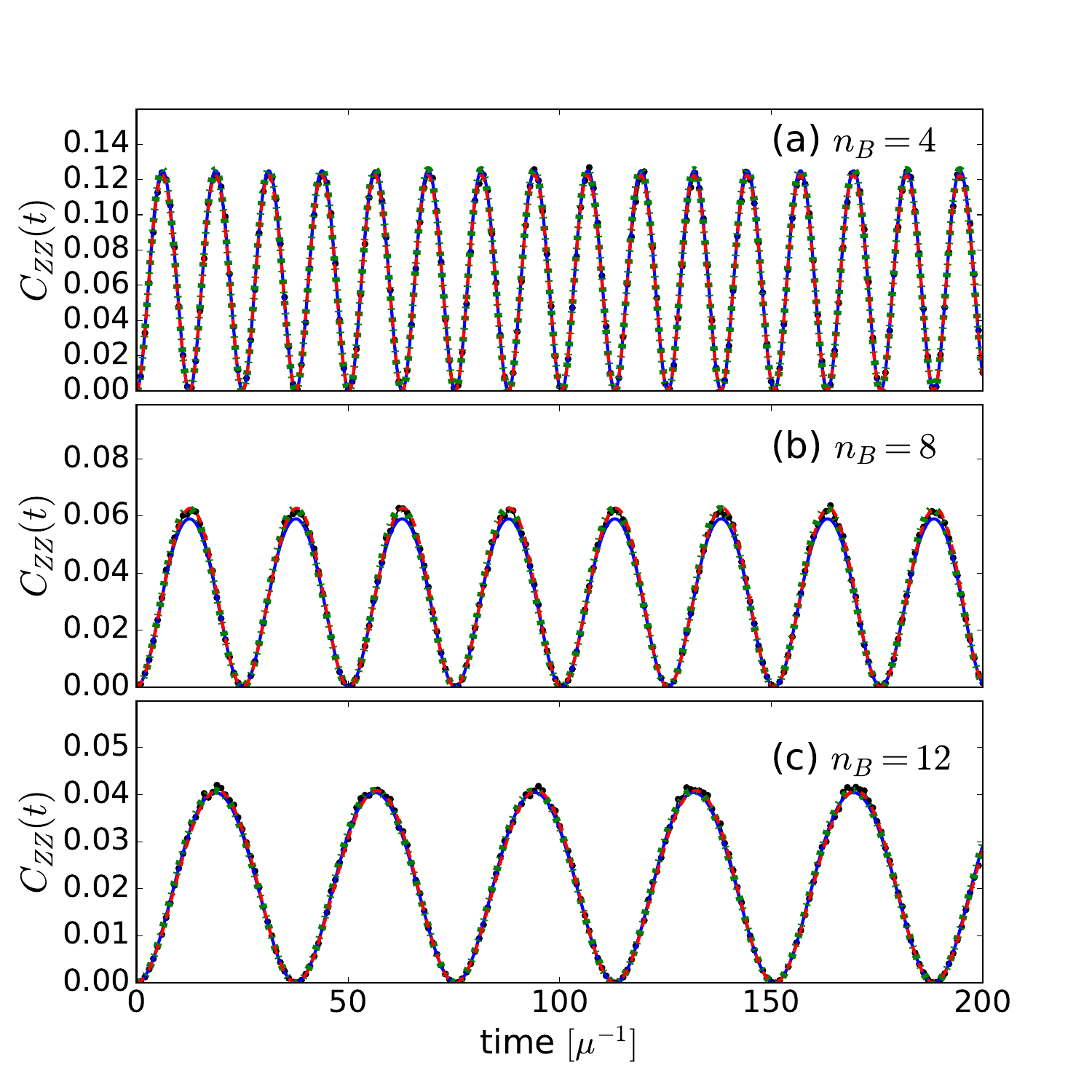} 
    \caption{Evolution of the average fluctuations $C_{ZZ} (t)$ given by Eq. (\ref{eq:total-zz}), for (a) $n_B=4$, (b) $n_B=8$, and (c) $n_B=12$ neutrinos. In all cases, half of the neutrinos are assumed to be in one of the states of the flavor basis and the other half in the other flavor state at initial time. The exact evolution (black line-filled circle), Gaussian sampling (blue solid line),
    bi-valued (red solid line) and Husimi sampling (green dotted line) are systematically shown. 
    }
    \label{fig:sigmaN4-8-12}
\end{figure}

We show in Fig. \ref{fig:zevolution} the evolution of the $z$-component of the operator in the flavor basis that was shown in Ref. \cite{Ami23} for small $n_B$ values, 
and that is related to the occupation of the two flavor states as illustrated by the first two equations in (\ref{eq:densityone}). We systematically compare this figure
with the PSA approach using the three sampling methods introduced previously. Despite the difference in initial polarization component values displayed in Fig. \ref{fig:sampling}, the three sampling methods give rather close results. The PSA method can qualitatively reproduce the evolution up to 60 $\mu^{-1}$ and then tends to overestimate the damping at a longer time. 
This overestimation is a known feature of the PSA \cite{Lac12}.  It should also be
kept in mind that (i) the semiclassical mapping made in the PSA approach is supposed to be more and more valid when the number of particles increases. From that point of view, $n_B=8$ is a rather extreme case; (ii) Although some differences can be seen in individual particle properties, collective properties obtained by averaging over particles might still be rather well reproduced. This will be illustrated below. 

 \begin{figure}[htbp] 
\includegraphics[width= \linewidth]{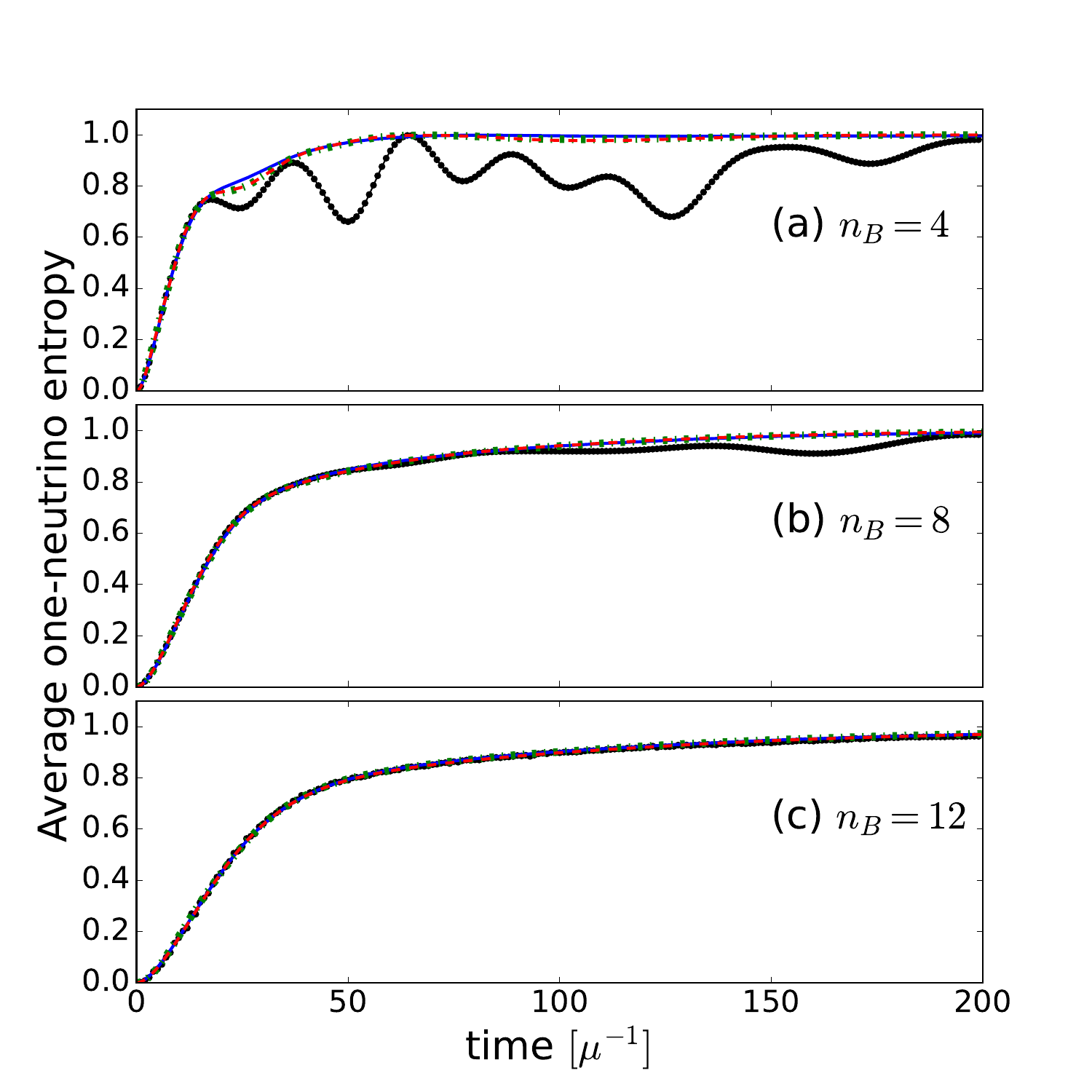} 
    \caption{Average one neutrino entropy obtained for $n_B=4$, $n_B=8$ and $n_B=12$ neutrino's beam corresponding to Fig. \ref{fig:sigmaN4-8-12}. The conventions for the curves are the same as in 
    Fig. \ref{fig:sigmaN4-8-12}.  }
    \label{fig:EntropN4-8-12}
\end{figure}

Item (i) was one of the motivations for improving the sampling method for small $n_B$ values. A careful analysis of Fig. \ref{fig:zevolution} 
shows that the Husimi and bi-valued technique that does not pre-suppose a Gaussian probability distribution is slightly better/less damped than the Gaussian sampling when focusing on time $t \le 60 \mu^{-1}$. Still, all methods lead to very close results in general, showing the robustness of the approach. Note that a pure mean-field approximation would fail to reproduce these evolutions. A second remarkable feature is that, despite the differences observed on individual properties of neutrinos in Fig. \ref{fig:zevolution}, the PSA is highly predictive for collective observables, i.e., observables obtained by summing over individual observables. Two examples are given below: quantum fluctuations and the average one neutrino entropy.

To illustrate the average properties of neutrinos, the average fluctuations of the $Z_\alpha$ quantities, defined as 
\begin{eqnarray}
C_{ZZ}(t) &=& \frac{1}{n_B^2} \sum_{\alpha,\beta} \left[ \langle Z_\alpha Z_\beta \rangle  - \langle Z_\alpha \rangle \langle Z_\beta \rangle \right], \label{eq:total-zz} 
\end{eqnarray}  
is shown as a function of time for an increasing number of beams in Fig. \ref{fig:sigmaN4-8-12}. We see that all sampling methods, even for the smallest $n_B$, are able to reproduce the exact solutions perfectly. Noteworthy, this quantity is linked to two-body effects badly accounted for in the standard mean-field approach. It is quite remarkable to observe that the fluctuation evolution are almost independent of the sampling method provided that the sampling leads to the proper first and second moment of the initial conditions. This underlines the robustness of the method to predict effects beyond the mean-field approximation.  

Dissipation and entanglement due to the interaction between neutrinos are other important aspects of neutrino physics.  
In \cite{Lac22}, the PSA was shown to quantitatively describe one- and two-body entropies for two beams with many particles in the beam. 
This is confirmed here for many beams and one particles per beam. We show in Fig. \ref{fig:EntropN4-8-12} the average one-neutrino entropy 
for an increasing number of beams with one neutrino per beam.  This quantity is obtained from the equation:  
\begin{eqnarray}
S &=& - \frac{1}{n_B} \sum_\alpha \left[ \lambda^\alpha_0 (t) \log_2 \lambda^\alpha_0 (t)+ \lambda^\alpha_1 (t) \log_2  \lambda^\alpha_1 (t) \right] ,
\end{eqnarray}   
where $(\lambda^\alpha_0(t), \lambda^\alpha_1(t))$ are the two eigenvalues of $\rho^\alpha$. Since all one-body densities are rank 2 matrices, we have the upper bound $S \le 1$. 
We see in Fig. \ref{fig:EntropN4-8-12} that PSA results reproduce rather well the exact evolution of this quantity. For $n_B=4$, small deviations between 
the exact and approximate treatments could be attributed to the deviations already seen in Fig. \ref{fig:zevolution}. Still, a focus on time $t \le 60 \mu^{-1}$ shows 
that the bi-valued method is slightly better than the two other sampling techniques.  When $n_B$ increases, the PSA results become more predictive, and
for $n_B=12$, PSA results cannot be distinguished from the exact evolution. In all cases, PSA gives the upper envelope limits for the average one-neutrino entropy.  

\begin{figure}[htbp] 
\includegraphics[width=\linewidth]{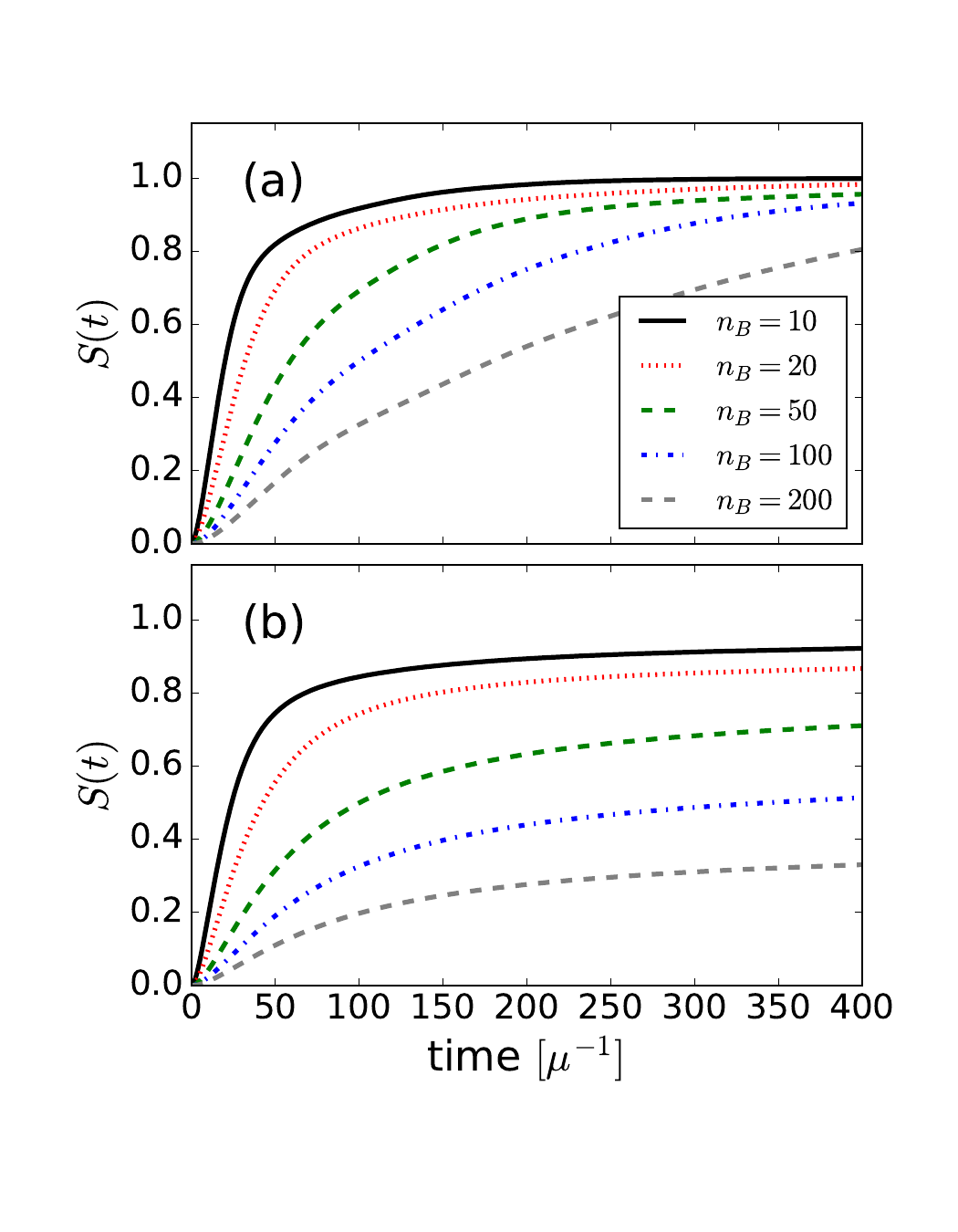} 
    \caption{Illustration of the average one neutrino entropy increase as a function of time obtained with the phase-space method for various numbers of beams $n_B$. The case of one neutrino per beam is considered in all cases, and the bi-valued sampling method is used. In panel (a) (resp. panel (b))
    the case of time-independent (resp. time-dependent) coupling between neutrinos is displayed. For the time-dependent case, the interaction is given by Eq. (\ref{eq:paramtime}).}
    \label{fig:entropyincrease}
\end{figure}

The conclusion of the present example is that, as soon as the number of beams increases and exceeds $10$, the PSA approach has very high predictive power for the Hamiltonian case of Ref. \cite{Ami23}. 
In particular, it reproduces many aspects of the exact evolution 
at a much lower numerical cost. While the exact solution can hardly be 
made for $n_B$ greater than $20$, we illustrate the evolution 
of the entropy obtained up to $n_B=200$, for one neutrino per beam 
in Fig. \ref{fig:entropyincrease}. These results have been obtained on a standard desk computer. 

Below, we give further validations of PSA for time-dependent Hamiltonian that are relevant for the neutrino problems, as well for increasing
number of neutrinos.           

\subsection{Application to three beams with many neutrinos}
\label{sec:3neutrinos}

Our previous study demonstrated that PSA can simulate the evolution of many interacting beams with one neutrino in each beam. One attractive aspect of the PSA technique is its numerical scaling when arbitrarily increasing the number of neutrinos in each beam. Specifically, the number of equations of motion to be solved for $n_B$ beams is $3 n_B$, independently from the numbers of neutrinos in different beams. For a given beam, the number of random numbers to generate might depend on 
the neutrino number if we use bi-valued sampling but is independent of 
the number of neutrinos in the beam in the two other methods. Noteworthy, the three sampling methods lead to identical results 
as soon as the number of neutrinos in the beam exceeds a few. Here, we illustrate the predictive power of the PSA with many neutrinos in each beam. We have already shown in Ref. \cite{Lac22}, by comparing to the exact results of Ref. \cite{Mar21} for $2$ beams with $50$ or more neutrinos per beam, that the PSA reproduced efficiently this exact evolution. 

As far as we know, the only attempt to perform an exact simulation with relatively large numbers of neutrinos per beam
was in Ref. \cite{Rog22a}, and is restricted to $n_B=3$. In this case, the three beams interact through the Hamiltonian
\begin{eqnarray}
H &=& \sum_{\alpha \neq \beta}^{n_B} G_{\alpha \beta}(t) \vec{J}_{\alpha} \cdot \vec{J}_{\beta}, \label{eq:hamil3beams}
\end{eqnarray} 
with:
\begin{eqnarray}
G_{12} &=& G_{21} = \frac{2}{N} , \nonumber \\
G_{13} &=& G_{31} = \frac{1}{N} (1-c) , \nonumber \\
G_{23} &=& G_{32} = \frac{1}{N} (1+c). \nonumber 
\end{eqnarray} 
The initial state is assumed to be 
\begin{eqnarray}
| \Psi (0) \rangle &=& | 1^A_f \rangle^{\otimes N_A} \otimes  |0^B_f \rangle^{\otimes N_B} \otimes  | 1^C_f  \rangle^{\otimes N_C}
\end{eqnarray}
with $N_A$, $N_B$ and $N_C$ the number of neutrinos in the beam $A$, $B$, and $C$ respectively, and $N=N_A+N_B+N_C$. $(| 0^{A/B,C}_f \rangle , | 1^{A/B,C}_f \rangle )$  are schematic notations for the flavor states in the three beams. 
 \begin{figure}[htbp] 
\includegraphics[width=\linewidth]{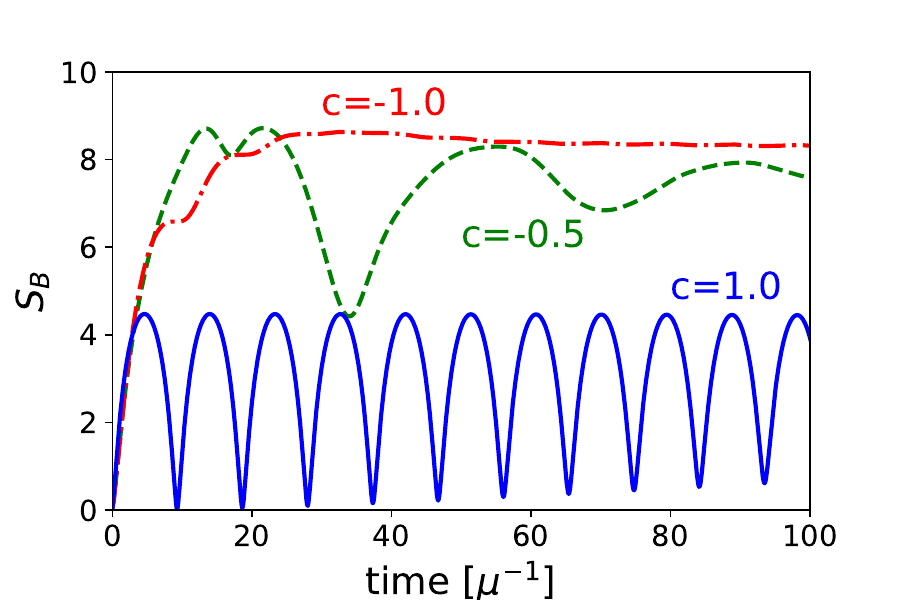} 
    \caption{Many-body entropy estimated for the system $B$ 
    for the three beams interacting through the Hamiltonian (\ref{eq:hamil3beams}) for three different values of the $c$ parameters. The approximate entanglement entropy is obtained using the expressions (\ref{eq:entropapp}) and (\ref{eq:gammaalpha}). This figure reproduces the results obtained in Fig. 12 of Ref. \cite{Rog22a}.
    }
    \label{fig:testRoggero2}
\end{figure}

In Ref. \cite{Rog22a}, the exact evolution of the entropy of one of the beams $\alpha \in \{A,B,C\}$ was obtained. This could be done by first obtaining the reduced entropy of this beam through a partial trace of the full density over the other beams. This reduced density can then be 
diagonalized to compute the entanglement entropy. 
An exact estimate of the entropy in PSA is a priori possible. Still, it would require first building the reduced density matrix of the beam, which becomes rapidly intractable when the number of neutrinos increases as this reduced density matrix has a size $2^{N_\alpha}$. An illustration of a strategy that can be used in PSA to obtain the two-neutrino entropy can be found in Ref. \cite{Lac22}. As $N_\alpha$ increases, the brute-force approach becomes difficult. Alternatively, 
one can follow the method proposed in Ref. \cite{Rog22a}, where the entropy is approximately computed from the second moment of the polarization vector. Using the PSA method, where the quantum 
fluctuations are replaced by classical average at all times, the
entropy of the beam $\alpha$, denoted by $S_\alpha(t)$, can be estimated from the expression:
\begin{eqnarray}
    S_\alpha(t) &=& \frac{ 1 + 2 \Gamma_\alpha(t)}{\log(2)} {\rm arccoth}\left(1 + 2 \Gamma_\alpha(t) \right) \nonumber \\
    &+& \frac{1}{2} \left[ \log_2 \Gamma_\alpha(t) +\log_2 \left(
    1 + \Gamma_\alpha (t) \right)  \right]. \label{eq:entropapp}
\end{eqnarray}
Within the PSA approach, the quantity $\Gamma_{\alpha}(t)$ is estimated from the phase-space average:
\begin{eqnarray}
\Gamma_{\alpha}(t) = \frac{N_\alpha}{4} \left[ 1 -  \overline{P^{\alpha (\lambda)}_z(t)  P^{\alpha (\lambda)}_z (t)}  \right]. \label{eq:gammaalpha}
\end{eqnarray}  
We show in Fig. \ref{fig:testRoggero2}, an illustration of results obtained with some sets of parameters tested in Ref. \cite{Rog22a} and 
should be compared with the Fig. 12 of the same reference.  
The results obtained with the PSA reproduces quantitatively very well 
those obtained using the exact estimate of the fluctuations.

\subsubsection{Applications to time-dependent Hamiltonian case}
\label{sec:timedependent}

Following Ref. \cite{Ami23} and with the objective to reproduce the results presented in this work, we have previously considered the unrealistic situation where the coupling between neutrinos is kept constant in time. When emitted from a stellar object, the coupling being proportional to the relative angles in forward direction 
is expected to decrease as the distance from the emitter increases. To account for this effect, one can generalize Eq. (\ref{eq:coupling_matrix}) by introducing the time-dependent coupling:
\begin{eqnarray}
    G_{\alpha,\beta}&=&\frac{\mu}{N}\left(1-\cos\left[ \Omega_{\rm max} (t)\frac{|\alpha-\beta|}{n_B-1}\right] \right).
    \label{eq:paramtime}
\end{eqnarray}
$\Omega_{\rm max} (t)$ is defined as a function of the radius of the emitter denoted by $R_\nu$
and $r(t)$ the distance from the emitter:
\begin{eqnarray}
    \Omega_{\rm max} (t) &=& 2 {\rm arcsin} \left[ \frac{R_\nu}{r(t)} \right]. \label{eq:omegat}
\end{eqnarray}
Here we assume $R_\nu = 32.2 \mu^{-1}$ as in \cite{Pat21}. Note that even if we set $R_\nu$ to the same value
as in this reference, the time-dependent coupling is slightly different from the one used in \cite{Pat21} and consists of a direct generalization of the time-independent case considered in \cite{Ami23}. Specifically, it accounts for the fact that the maximal relative angles $\Omega_{\rm max} (t)$ between two neutrinos decrease in time after 
emission from the same stellar source and assumes that the relative angles between different beams are obtained 
by a uniform discretization between $0$ and $\Omega_{\rm max} (t)$. The time-independent calculations
considered previously assume $\Omega_{\rm max} (t) = \Omega_{\rm max} (0) = \arccos(0.9)$ which corresponds to start the calculation at distance $r(0) \equiv r_0 = 144 \mu^{-1}$  and keep the coupling 
unchanged during the evolution.

 \begin{figure}[htbp] 
\includegraphics[width=\linewidth]{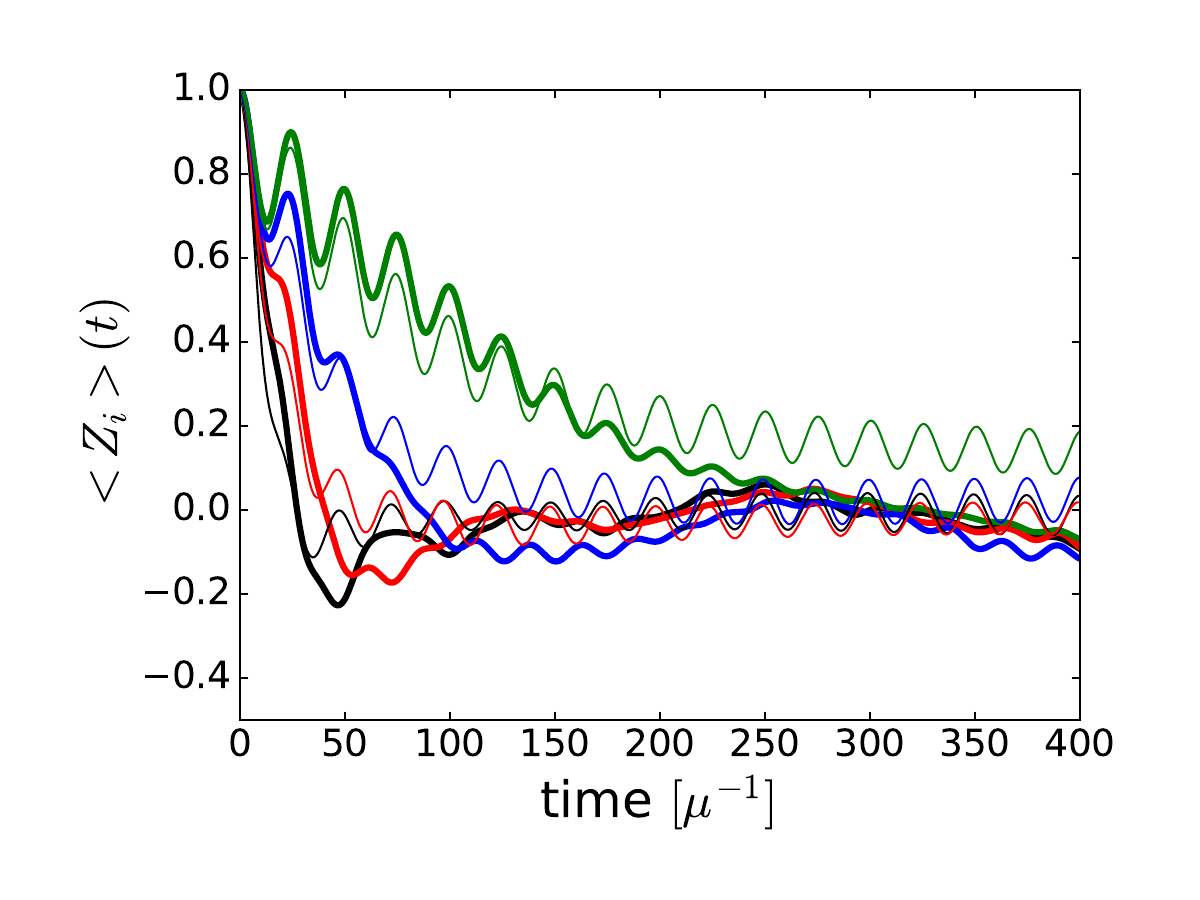} 
    \caption{Evolution of the $\langle {\cal Z}_\alpha \rangle$ components obtained with the PSA for each neutrino with the same initial condition as in Fig. \ref{fig:zevolution} with $n_B=8$ and one neutrino per beams. The reference 
    result of Fig. \ref{fig:zevolution}(c) for time-independent neutrino-neutrino interaction are shown with thick lines, while the results obtained with time-dependent coupling using Eq. (\ref{eq:paramtime}) are shown with thin lines. In both cases, the bi-valued sampling technique was used.} 
    \label{fig:ztime-td}
\end{figure}

To account for the reduction of the coupling, we assume below that the initial distance is still $r_0$, but now $r(t) = r_0 + t$ in Eq. (\ref{eq:omegat}), where $t$ is given in $[\mu^{-1}]$ units. 
We illustrate in Fig. \ref{fig:ztime-td} the effect of this reduction on the individual properties of neutrinos. In this example, $n_B=8$ neutrino beams are considered with one neutrino per beam, as in Fig. \ref{fig:zevolution}. We observe in this figure that, at the early stage of the evolution, 
the damping of the $\langle {\cal Z}_\alpha \rangle$ evolutions is similar to the time-independent case. Still, for longer times, a significant difference is seen between the time-independent and time-dependent case. In particular, the time-dependent coupling case has less beating and tends to stabilize towards harmonic oscillations. Note surprisingly, the damping effect induced by the coupling between beams is less pronounced. This is further illustrated in Fig. \ref{fig:entropyincrease}(b), where the average entropy is shown and compared to 
the time-independent case (panel (a)). In the time-independent coupling case, provided that the system is evolved for a sufficient time that increases with $n_B$, the average one-neutrino entropy is always reaching its maximum value $S_{\rm max} = 1$ asymptotically. When time-dependent coupling is considered, we see that the entropy saturates at a value lower than $1$. This indicates that the interaction vanishes before reaching the maximum disorder accessible to the full system. In parallel, we also expect that 
the reduction of interaction will quench the entanglement between different neutrino beams. Notably, this quenching and/or absence of reaching the maximum entropy will significantly depend 
on the initial time (or distance $r_0$) used. Future studies with the phase-space approach might allow us to investigate this dependence systematically. In particular, $r_0$ is sometimes taken large enough
so that adiabatic approximation holds, allowing the exact evolution to be performed when $n_B$ is not very large. The phase-space method does not invoke adiabaticity and can be used in the non-adiabatic regime.  

\subsection{Increasing the number of neutrinos and impact of the initial conditions} 

 \begin{figure}[htbp] 
\includegraphics[width=\linewidth]{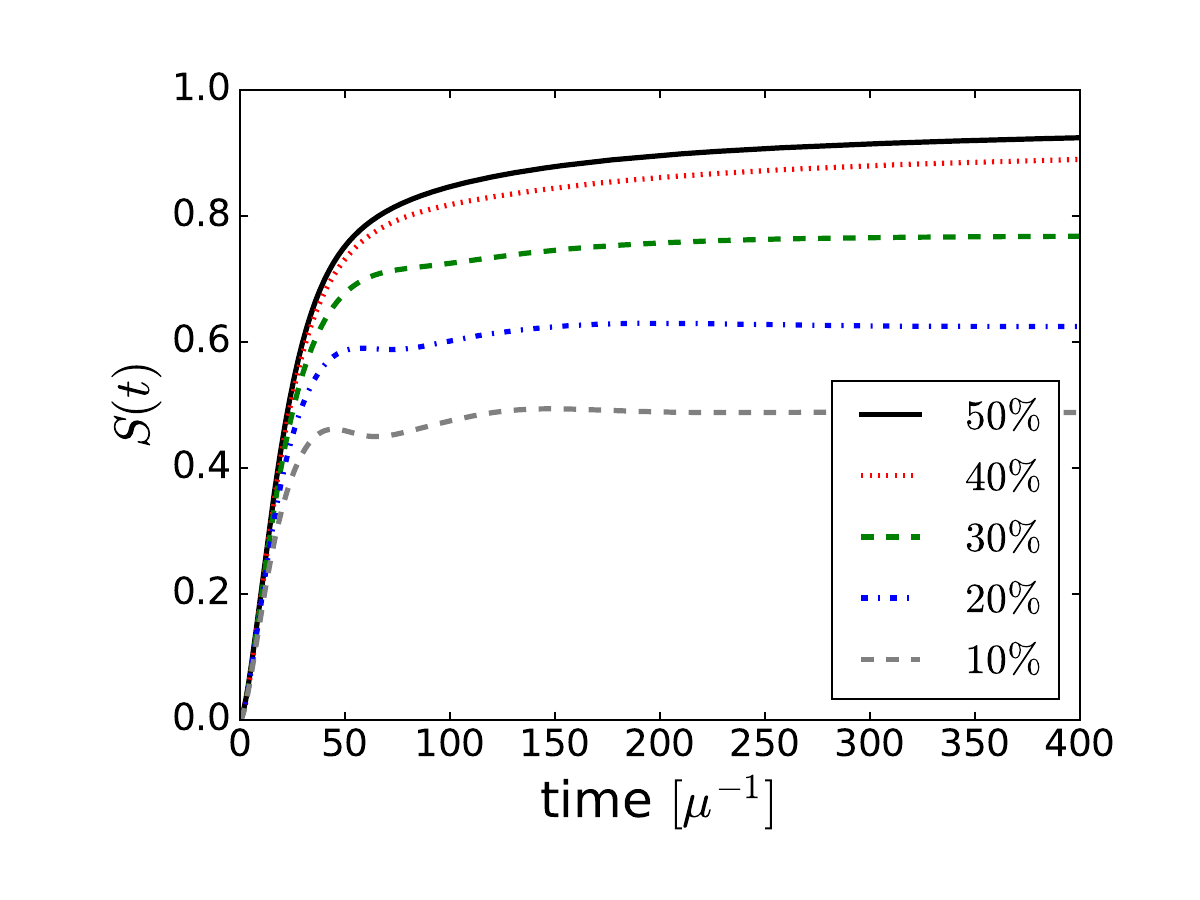} 
    \caption{One neutrino average entropy evolution obtained for $n_B=10$ coupled neutrinos beams with the time-dependent coupling given by Eq. (\ref{eq:paramtime}). 
    initially, the first $i=1, \cdots, n_B/2$ (resp. last $i =n_B/2+1, \cdots, n_B $) beams are initialized with $n_0 = 2$, $4$, $6$, $8$ and $10$ (resp. 
    $n_1 = 18$, $16$, $14$, $12$ and $10$) in the state $|0^f_{\alpha}\rangle$
    (resp. $|1^f_{\alpha}\rangle$), so that the total number of neutrinos is $N=100$ in all simulations. The different $(n_0,n_1)$ values corresponds respectively to $10 \%$, $\cdots$, $50 \%$ of the total number of neutrinos in the state $|0^f_{\alpha}\rangle$.  }
    \label{fig:entropyInbalanced}
\end{figure}

Contrary to the exact solution, increasing the number of particles in each beam does not lead to additional numerical cost since the number 
of equations to solve remains $3 n_B$, whatever the number of neutrinos per beam. When using the Hamiltonian (\ref{eq:Hamiltonian}) where the 
couplings are normalized to the total number of neutrinos $N$, it is easy to realize that the phase-space method results when applied with the Gaussian sampling method 
will be unchanged if one increases $N$ while keeping all ratios $\{r_\alpha = N_\alpha/N \}$ unchanged. This stems from the fact that (i) 
in the mean-field equation of motion given by Eq. (\ref{eq:eom-mf}), only these ratios appear once using the expressions (\ref{eq:coupling_matrix}) for the $G_{\alpha,\beta}$, and (ii) the Gaussian 
sampling given by Eq. (\ref{eq:gaussampling}) leads to a Gaussian sampling on the polarization vector components that is independent of the $N_\alpha$ values. Said differently, the numerical cost 
to consider  $1$ or $10$ million neutrinos for each beam will be unchanged since in both cases $N_\alpha / N = {n_B}^{-1}$ for both cases. In the case of bi-valued distribution, only the properties (i) hold, but due to the central limit theorem, as we illustrated already, the evolution becomes rapidly independent of the sampling method as soon as more than $N=10$ particles are considered, i.e. as soon as the total number of particles increases.        

Similarly to what was done in section \ref{sec:3neutrinos} when changing the $c$ parameters in the coupling constants, one can change the 
different ratios $\{ r_\alpha\}$ while keeping the total number of particles fixed. 

In the simulation shown previously for the time-dependent Hamiltonian, we always assumed that the beams separate 
into two sets, one set of beams where all neutrinos are initially in one of the flavor states while the other set of beams are in the other flavor state 
at the initial time. We also assumed that all beams have the same number of neutrinos. We now consider a different situation where we have an imbalanced population between the two initial flavor states. Specifically, we still consider that the 
$n_B/2$ first beams all have neutrinos initialized in the $|0^f_{\alpha}\rangle$ and the other beams have neutrinos 
initialized in the $|1^f_{\alpha}\rangle$ state. But now we assume that each beam in the first set of beams contain 
$n_0$ neutrinos while the others contains $n_1$ neutrinos. Accordingly, we have the ratios:
\begin{eqnarray}
    \{ r_\alpha \}_{\alpha=1, \cdots, n_B/2} = \frac{2 n_0} {n_B(n_0+n_1)} , \nonumber \\
    \{ r_\alpha \}_{\alpha=n_B/2+ 1, \cdots, n_B} = \frac{2 n_1} {n_B(n_0+n_1)}  \nonumber
\end{eqnarray} 
In the illustration shown below, we fix both the number of beams $n_B$ and 
the total number of particles $N = \frac{n_B}{2} (n_0 + n_1)$ and vary the ratio $n_0/n_1$. Specifically, we consider the case of $n_B=10$ beams and vary the number of particles such that we have $ n_0 n_B/2 = 10$, $\cdots$, $50$ that is equivalent
to having $10 \%$, $\cdots$, $50 \%$ of the total number of neutrinos initially in the state $|0^f_{\alpha}\rangle$. Note that 
due to the symmetry of the problem with the exchange between $|0^f_{\alpha}\rangle$ and $|1^f_{\alpha}\rangle$ higher percentage can be deduced from the one simulated here. Results corresponding to different percentages of states initially in different flavor states are shown in Fig. \ref{fig:entropyInbalanced}. We observe 
in this figure that the non-equilibrium entropy is significantly affected by the initial population of the $|0^f_{\alpha}\rangle$ compared to the 
$|1^f_{\alpha}\rangle$ state population. Changing the percentage compared to an equal initial population leads to quenching the asymptotic entropy. An equal 
population gives the absolute maximum entropy that can be reached. This behavior is what one would expect based on previous calculations on unbalanced initial states but time-independent Hamiltonians 
\cite{Fri06,Mar21}

This illustrative example points out the significant effect of the initial conditions on the entanglement of neutrinos. Most of the many-body applications 
performed today assume rather simple initial states. The sensitivity to initial conditions observed in Fig. \ref{fig:entropyInbalanced} clearly indicates the necessity to perform future applications with more realistic initial states.  

\section{Conclusion}

In this work, the phase-space approach proposed recently to simulate neutrino oscillation \cite{Lac22}
is further illustrated and benchmarked. To treat neutrino beams with several neutrinos varying from one to millions of neutrinos, a novel statistical sampling method is used to prepare a set of initial conditions used later for the time evolution. This method is compared to existing sampling methods based on Gaussian probability assumption or Husimi quasi-probabilities. We show that the new method improves the reproduction of the evolution when very few particles are used. The three sampling methods give identical results when the total number of particles exceeds ten. 

The phase-space approach is confronted with exact results obtained when available. We show that in all cases, it is possible at a much lower numerical cost to reproduce these exact results well. Only small deviations are seen in the limit where very few neutrinos are considered. The capacity of the phase-space approach to treat a large number of beams with a large number of neutrinos or to treat the problem of realistic time-dependent couplings between neutrinos is illustrated 
here. In particular, we identified several sources of quenching of the average one-neutrino entropy 
either induced by the time-dependence of the neutrino-neutrino interaction and/or by the different population of each neutrino flavor state at initial time. In both cases, we argue that a precise description of realistic initial conditions will be necessary to achieve meaningful conclusions on the neutrino oscillations or entanglement patterns.       

We concentrate here on the proof that the PSA approach can be highly predictive for the neutrino oscillation problem and can be used as an alternative to the exact solution when this last solution 
is undoable. It will, for instance, be a useful tool to compare simulations made with quantum computers 
when these technologies will surpass classical computers. 

The phase-space approach is applied to rather simple initial condition where the system is prepared in a pure state. However, it can easily be extended to treat the emission from a thermal statistical ensemble \cite{Ayi08,Lac14}. We also anticipate that it could be improved by accounting for the MSW effect and/or by going beyond the SU(2) approximation.

\section{Acknowledgments }

This project has received financial support from the CNRS through the AIQI-IN2P3 project.  This work is part of 
HQI initiative (\href{www.hqi.fr}{www.hqi.fr}) and is supported by France 2030 under the French 
National Research Agency award number ``ANR-22-PNQC-0002''. A.B.B. was supported by U.S. Department of Energy, Office of Science, Office of High Energy Physics, under Award No. DE-SC0019465 and by US National Science Foundation award No. PHY-2411495. We acknowledge the use of IBM Q cloud as well as the use of the Qiskit software package
\cite{qiskit2024} for performing the quantum simulations. D.L. and A.R. also thank the QC4HEP Working Group for discussions.


\begin{thebibliography}{99}



\bibitem{Dua06} H. Duan, G. M. Fuller, J. Carlson, and Y.-Z. Qian, {\it Simulation of coherent nonlinear neutrino flavor transformation in the supernova environment: Correlated neutrino trajectories}, Phys. Rev. {\bf D 74}, 105014 (2006).

\bibitem{Bah07}  A B Balantekin and Y Pehlivan, {\it Neutrino–neutrino interactions and flavor mixing in dense matter},  J. Phys. {\bf G 34}, 47 (2007).

\bibitem{Vol24} M. Cristina Volpe, {\it Neutrinos from dense environments: Flavor mechanisms, theoretical approaches, observations, and new directions}, 
Rev. Mod. Phys. 96, 025004 (2024)


\bibitem{Wol78} L. Wolfenstein, {\it Neutrino oscillations in matter}, Phys. Rev. D 17, 2369 (1978). 
\bibitem{Mik85} S. P. Mikheyev et A. Yu. Smirnov, {\it Resonance amplification of oscillations in matter and spectroscopy of solar neutrinos}, 
Soviet Journal of Nuclear Physics, {\bf 42}, 913 (1985). 



\bibitem{Ful87} G. M. Fuller, R. W. Mayle, J. R. Wilson, and D. N.
Schramm, {\it Resonant Neutrino Oscillations and Stellar Collapse}, Astrophys. J. {\bf 322}, 795 (1987).

\bibitem{Not88} D. Notzold and G. Raffelt, {\it Neutrino dispersion at finite temperature and density}, Nucl. Phys. {\bf B 307}, 924 (1988).
\bibitem{Sig93} G. Sigl and G. Raffelt, {\it General kinetic description of relativistic mixed neutrinos},  Nucl. Phys. B 406, 423 (1993).



\bibitem{Peh11}
Y. Pehlivan, A. B. Balantekin, Toshitaka Kajino, and Takashi Yoshida, {\it Invariants of collective neutrino oscillations},  
Phys. Rev. {\bf D 84}, 065008 (2011). 




\bibitem{Bir18} Savas Birol, Y. Pehlivan, A. B. Balantekin, and T. Kajino, {\it Neutrino spectral split in the exact many-body formalism},
Phys. Rev. {\bf D 98}, 083002 (2018). 

\bibitem{Pat19} Amol V. Patwardhan, Michael J. Cervia, and A. Baha Balantekin, {\it Eigenvalues and eigenstates of the many-body collective neutrino oscillation problem}, Phys. Rev. {\bf D 99}, 123013 (2019). 


\bibitem{Rra19} Ermal Rrapaj, {\it Exact solution of multi-angle quantum many-body collective neutrino flavor oscillations}, Phys. Rev. {\bf C 101}, 065805 (2020).


\bibitem{Mar21}  Joshua D. Martin, A. Roggero, Huaiyu Duan, J. Carlson, V. Cirigliano, {\it Classical and Quantum Evolution in a Simple Coherent Neutrino Problem}, Phys. Rev. {\bf D 105}, 083020 (2022).

\bibitem{Xio22} Zewei Xiong, {\it Many-body effects of collective neutrino oscillations}, Phys. Rev. D 105, 103002 (2022). 


\bibitem{Ill22} Marc Illa, Martin J. Savage, {\it Multi-Neutrino Entanglement and Correlations in Dense Neutrino Systems},  	Phys. Rev. Lett. {\bf 130}, 221003 (2023).

\bibitem{Lac22} Denis Lacroix, A. B. Balantekin, Michael J. Cervia, Amol V. Patwardhan, and Pooja Siwach, {\it Role of non-Gaussian quantum fluctuations in neutrino entanglement},  Phys. Rev. {\bf D 106}, 123006 (2022). 


\bibitem{Cer22}  Michael J. Cervia, Pooja Siwach, Amol V. Patwardhan, A. B. Balantekin, S. N. Coppersmith, Calvin W. Johnson, 
 {\it Collective neutrino oscillations with tensor networks using a time-dependent variational principle}, Phys. Rev. D 105, 123025 (2022).  
 


\bibitem{Rog22a} Alessandro Roggero, Ermal Rrapaj, and Zewei Xiong, {\it Entanglement and correlations in fast collective neutrino flavor oscillations}, 
Phys. Rev. {\bf D 106}, 043022 (2022).

\bibitem{Mar23a}  Joshua D. Martin, A. Roggero, Huaiyu Duan, J. Carlson, {\it Many-body neutrino flavor entanglement in a simple dynamic model }, arXiv:2301.07049

\bibitem{Mar23b} Joshua D. Martin, Duff Neill, A. Roggero, Huaiyu Duan, and J. Carlson, {\it Equilibration of quantum many-body fast neutrino flavor oscillations}, 
Phys. Rev. D 108, 123010 (2023).


\bibitem{Bha23} Ramya Bhaskar, Alessandro Roggero, Martin J. Savage, {\it Time Scales in Many-Body Fast Neutrino Flavor Conversion}, arXiv:2312.16212. 



\bibitem{Cer19} Michael J. Cervia, Amol V. Patwardhan, A.B. Balantekin, S.N. Coppersmith, and Calvin W. Johnson,
 {\it Entanglement and collective flavor oscillations in a dense neutrino gas}, Phys. Rev. {\bf D 100}, 083001 (2019). 
 

\bibitem{Pat21} Amol V. Patwardhan, Michael J. Cervia, and A.B. Balantekin, {\it Spectral splits and entanglement entropy in collective neutrino oscillations}, Phys. Rev. {\bf D 104}, 123035 (2021).  



 \bibitem{Bal22}  A.B. Balantekin, {\it Quantum Entanglement and Neutrino Many-Body Systems}
J. Phys.: Conf. Ser. 2191 012004 (2022). 
 
 \bibitem{Rog21}  Alessandro Roggero, {\it Entanglement and Many-Body effects in Collective Neutrino Oscillations }, Phys. Rev. {\bf D 104}, 103016 (2021).


 
\bibitem{Hal21} Benjamin Hall, Alessandro Roggero, Alessandro Baroni, and Joseph Carlson, {\it Simulation of collective neutrino oscillations on a quantum computer}, Phys. Rev. {\bf D 104}, 063009. 

\bibitem{Yet22}  K\"ubra Yeter-Aydeniz, Shikha Bangar, George Siopsis, and Raphael C. Pooser, {\it Collective neutrino oscillations on a quantum computer}, Quantum Inf Process {\bf 21}, 84 (2022). 

\bibitem{Kum22} Abhishek Kumar Jha and Akshay Chatla, {\it Quantum studies of neutrinos on IBMQ processors}, Eur. Phys. J. Spec. Top. {\bf 231}, 141 (2022). 

\bibitem{Ill22b} Marc Illa, Martin J. Savage, {\it Basic Elements for Simulations of Standard Model Physics with Quantum Annealers: Multigrid and Clock States}, arXiv:2202.12340 [quant-ph]

\bibitem{Ami23} Valentina Amitrano, Alessandro Roggero, Piero Luchi, Francesco Turro, Luca Vespucci, Francesco Pederiva, {\it Trapped-Ion Quantum Simulation of Collective Neutrino Oscillations}, Phis. Rev. {\bf D 107}, 023007 (2023). 

\bibitem{Tur24}  Francesco Turro, Ivan A. Chernyshev, Ramya Bhaskar, Marc Illa, 
{\it Qutrit and Qubit Circuits for Three-Flavor Collective Neutrino Oscillations}, 
arXiv:2407.13914.  



\bibitem{Sha23} Shashank Shalgar and Irene Tamborra, {\it Do we have enough evidence to invalidate the mean-field approximation adopted to model collective neutrino oscillations?}
Phys. Rev. {\bf D 107}, 123004 (2023)

\bibitem{Joh23} L. Johns, {\it Neutrino many-body correlations}, to appear in International Journal of Modern Physics A,  arxiv:2305.04916. 

\bibitem{Koh24} Anson Kost, Lucas Johns, and Huaiyu Duan, {\it Once-in-a-lifetime encounter models for neutrino media: From coherent oscillations to flavor equilibration},
Phys. Rev. {\bf D 109}, 103037 (2024). 

\bibitem{Cir24} Vincenzo Cirigliano, Srimoyee Sen, Yukari Yamauchi {\it Neutrino many-body flavor evolution: the full Hamiltonian}, arxiv:2404.16690 



\bibitem{Ayi08} S. Ayik, {\it A stochastic mean-field approach for nuclear dynamics}, Phys. Lett. {\bf B 658}, 174 (2008).

\bibitem{Lac12} Denis Lacroix, Sakir Ayik, and Bulent Yilmaz, {\it Symmetry breaking and fluctuations within stochastic mean-field dynamics: Importance of initial quantum fluctuations},  Phys. Rev. {\bf C 85}, 041602(R) (2012) 
     
\bibitem{Lac14} D. Lacroix and S. Ayik, {\it Stochastic quantum dynamics beyond mean field}, Eur. Phys. J. {\bf A 50}, 95 (2014). 

\bibitem{Lac13} Denis Lacroix, Danilo Gambacurta, and Sakir Ayik, {\it Quantal corrections to mean-field dynamics including pairing}, 
Phys. Rev. {\bf C 87}, 061302(R) (2013). 

\bibitem{Yil14} Bulent Yilmaz, Denis Lacroix, and Resul Curebal, {\it Importance of realistic phase-space representations of initial quantum fluctuations using the stochastic mean-field approach for fermions}, Phys. Rev. {\bf C 90}, 054617 (2014).

\bibitem{Lac14b} Denis Lacroix, S. Hermanns, C. M. Hinz, and M. Bonitz, {\it Ultrafast dynamics of finite Hubbard clusters: A stochastic mean-field approach}, Phys. Rev. {\bf B 90}, 125112 (2014).   

\bibitem{Lac16} D. Lacroix, Y. Tanimura, S. Ayik, B. Yilmaz, {\it A simplified BBGKY hierarchy for correlated fermions from a stochastic mean-field approach} Eur. Phys. J. {\bf A 52}, 94 (2016).
 
\bibitem{Ulg19} Ibrahim Ulgen, Bulent Yilmaz, and Denis Lacroix, {\it Impact of initial fluctuations on the dissipative dynamics of interacting Fermi systems: A model case study}, Phys. Rev. {\bf C 100}, 054603 (2019). 


\bibitem{Reg18} David Regnier, Denis Lacroix, Guillaume Scamps, and Yukio Hashimoto, {\it Microscopic description of pair transfer between two superfluid Fermi systems: Combining phase-space averaging and combinatorial techniques}, 
Phys. Rev. C 97, 034627 (2018). 


\bibitem{Czu20} Thomas Czuba, Denis Lacroix, David Regnier, Ibrahim Ulgen  and Bulent Yilmaz, 
{\it Combining phase-space and time-dependent reduced density matrix approach to describe the dynamics of interacting fermions}, 
Eur. Phys. J. {\bf A 56}, 111 (2020). 

   
\bibitem{Cep95} D. M. Ceperley, {\it Path integrals in the theory of condensed helium}, 
Rev. Mod. Phys. {\bf 67}, 279 (1995)
\bibitem{Fou01} W. M. C. Foulkes, L. Mitas, R. J. Needs, and G. Rajagopal, {\it Quantum Monte Carlo simulations of solids}, 
Rev. Mod. Phys. {\bf 73}, 33 (2001)

\bibitem{Car15} J. Carlson, S. Gandolfi, F. Pederiva, S. C. Pieper, R. Schi-
avilla, K. E. Schmidt, and R. B. Wiringa, {\it Quantum Monte Carlo methods for nuclear physics} Rev. Mod.
Phys. {\bf 87}, 1067 (2015).


\bibitem{Bog46} N.N. Bogolyubov, {\it Kinetic equations}, J. Phys. (URSS) {\bf 10}, 256 (1946). 
\bibitem{Bor46} H. Born, H.S. Green, {\it A general kinetic theory of liquids I. The molecular distribution functions},  Proc. Roy. Soc. {\bf A 188}, 10 (1946).

\bibitem{Kir46} J.G. Kirwood, {\it The Statistical Mechanical Theory of Transport Processes I. General Theory}, J. Chem. Phys. {\bf 14}, 180 (1946).


\bibitem{Bon16} M. Bonitz, {\it Quantum Kinetic Theory} (Springer, Berlin, 2016). 






\bibitem{Bre02}  H.-P. Breuer and F. Petruccione, {\it The Theory of Open Quantum Systems}, (Oxford University Press, Oxford, 2002).


\bibitem{Gil76} R. Gilmore, {\it Q and P representations for spherical tensors}, J. Phys. A: Math. Gen. {\bf 9}, L65 (1976). 




\bibitem{Nie02} {M. A. Nielsen and I. L. Chuang, {\it Quantum information and quantum computation.}, Cambridge University Press (2010).}

\bibitem{Aar18} {S. Aaronson, {\it Shadow Tomography of Quantum States}, Proceedings of the 50th Annual ACM SIGACT Symposium on Theory of Computing 325 (2018). }

\bibitem{Hua20} {HY. Huang, R. Kueng, and J. Preskill, {\it Predicting many properties of a quantum system from very few measurements}, Nat. Phys. {\bf 16}, 1050 (2020).}

\bibitem{Rui24}  Edgar Andres Ruiz Guzman, Denis Lacroix, {\it Restoring symmetries in quantum computing using Classical Shadows }, Eur. J. Phys. {\bf A 60}, 112 (2024).



\bibitem{qiskit2024} Ali Javadi-Abhari,  Matthew  Treinish, Kevin Krsulich, Christopher J. Wood,  Jake Lishman, Julien Gacon, Simon Martiel, Paul D. Nation,
Lev S. Bishop, Andrew W.  Cross, Blake R.  Johnson, Jay M. Gambetta, {\it Quantum computing with {Q}iskit}, (2024), arXiv.2405.08810. 

\bibitem{Fri06} Alexander Friedland, Bruce H. J. McKellar, and Ivona Okuniewicz, {\it Construction and analysis of a simplified many-body neutrino model}, Phys. Rev. {\bf D 73}, 093002 (2006).

\end{thebibliography}
\end{document}